\documentclass[preprint,12pt]{elsarticle} 

\usepackage{color}
\usepackage{fullpage}
\usepackage{amsmath}
\usepackage{mathrsfs}
\usepackage{newfloat}
\usepackage{lineno}
\usepackage{subfig}
\usepackage{multirow}
\usepackage{caption}
\usepackage{pstool}
\usepackage{enumerate}
\usepackage{moreverb}
\biboptions{numbers,sort&compress}
\usepackage[colorlinks,bookmarksopen,bookmarksnumbered,citecolor=blue,urlcolor=blue]{hyperref}

\newcommand{\OR}[1]{\textcolor{red}{#1}}
\newcommand{\bs}[1]{{\boldsymbol{#1}}}
\newcommand{\ignore}[1]{}
\DeclareFloatingEnvironment{algorithm}

\usepackage{amssymb}

\makeatletter
\def\ps@pprintTitle{
 \let\@oddhead\@empty
 \let\@evenhead\@empty
 \def\@oddfoot{\footnotesize\itshape International Journal for Numerical Methods in Engineering\hfill \today}
 \let\@evenfoot\@oddfoot}
\makeatother

\begin{document}

\begin{frontmatter}



\title{An adaptive variational Quasicontinuum methodology for lattice networks with localized damage\tnoteref{t1}}


\author[affiliation1]{O. Roko\v{s}\corref{mycorrespondingauthor}}
\cortext[mycorrespondingauthor]{Corresponding author, presently at Department of Mechanical Engineering, Eindhoven University of Technology, P.O. Box~513, 5600~MB Eindhoven, The Netherlands.}
\ead{o.rokos@tue.nl}

\author[affiliation2]{R.H.J. Peerlings}

\author[affiliation1]{J. Zeman}

\author[affiliation3]{L.A.A. Beex}

\address[affiliation1]{Department of Mechanics, Faculty of Civil Engineering, Czech Technical University in Prague, Th\'{a}kurova~7, 166~29 Prague~6, Czech Republic.}

\address[affiliation2]{Department of Mechanical Engineering, Eindhoven University of Technology, P.O. Box~513, 5600~MB Eindhoven, The Netherlands.}

\address[affiliation3]{Facult\'{e} des Sciences, de la Technologie et de la Communication, Campus Kirchberg, Universit\'{e} du Luxembourg, 6~rue Richard Coudenhove-Kalergi, L-1359 Luxembourg.}

\tnotetext[t1]{This is the accepted version of the following article: O. Roko\v{s}, R.H.J. Peerlings, J. Zeman, and L.A.A. Beex, An adaptive variational Quasicontinuum methodology for lattice networks with localized damage, \emph{Int. J. Numer. Meth.}, 112(2):147--200, 2017, which has been published in final form at \OR{\href{http://onlinelibrary.wiley.com/doi/10.1002/nme.5518/abstract}{10.1002/nme.5518}}. This article may be used for non-commercial purposes in accordance with the \href{http://olabout.wiley.com/WileyCDA/Section/id-828039.html}{Wiley Self-Archiving Policy}.}

\begin{abstract}
Lattice networks with dissipative interactions can be used to describe the mechanics of discrete meso-structures of materials such as 3D-printed structures and foams. This contribution deals with the crack initiation and propagation in such materials and focuses on an adaptive multiscale approach that captures the spatially evolving fracture. Lattice networks naturally incorporate non-locality, large deformations, and dissipative mechanisms taking place inside fracture zones. Because the physically relevant length scales are significantly larger than those of individual interactions, discrete models are computationally expensive. The Quasicontinuum~(QC) method is a multiscale approach specifically constructed for discrete models. This method reduces the computational cost by fully resolving the underlying lattice only in regions of interest, while coarsening elsewhere. In this contribution, the (variational) QC is applied to damageable lattices for engineering-scale predictions. To deal with the spatially evolving fracture zone, an adaptive scheme is proposed. Implications induced by the adaptive procedure are discussed from the energy-consistency point of view, and theoretical considerations are demonstrated on two examples. The first one serves as a proof of concept, illustrates the consistency of the adaptive schemes, and presents errors in energies. The second one demonstrates the performance of the adaptive QC scheme for a more complex problem.
\end{abstract}


\end{frontmatter}



%
%
\section{Introduction}
\label{Sect:Introduction}
Lattice networks are frequently employed to describe the mechanical response of materials and structures that are discrete by nature at one or more length scales, such as 3D-printed structures, woven textiles, paper, or foams. For lattice networks representing fibrous microstructures for instance, individual fibres can be identified with one-dimensional springs or beams. Further examples are the models of e.g.~\cite{RidruejoFiberGlass,LiuPaper,KulachenkoPaper,BeexTextile}.

The reason why lattice models may be preferred over conventional continuum theories and Finite Element~(FE) approaches, is twofold. First, the meaning and significance of the physical parameters associated with individual interactions in the lattice networks is easy to understand, whereas the parameters in constitutive continuum models represent the small-scale mechanics only in a phenomenological manner. An example is the Young's modulus or ultimate strength of a spring or beam (fibre or yarn) versus that of the network. Second, the formulation and implementation of lattice models is generally significantly easier compared to that of alternative continuum models. Large deformations, large yarn re-orientations, and fracture are for instance easier to formulate and implement (cf. e.g. the continuum model of Peng and Cao~\cite{Peng:2005} that deals with large yarn re-orientations). Thanks to the simplicity and versatility of lattice networks, they are furthermore used for the description of heterogeneous cohesive-frictional materials such as concrete. The reason is that discrete models can realistically represent distributed microcracking with gradual softening, implement material structure with inhomogeneities, capture non-locality of damage processes, and reflect deterministic or stochastic size effects. Examples of the successful use of lattice models for such materials are given in~\cite{SCHLANGENconcrete,Cusatis:2006,GrasslConcreteLattice,Elias:2015}.

As lattice models are typically constructed at the meso-, micro-, or nano-scale, they require reduced-model techniques to allow for application-scale simulations. A prominent example is the Quasicontinuum~(QC) method, which specifically aims at discrete lattice models. The QC method was originally introduced for conservative atomistic systems by Tadmor \emph{et al.}~\cite{Tadmor:1996:QAD} and extended in numerous aspects later on, see e.g.~\cite{Curtin:2003:ACC,MillTad2002,Miller:2009:UFP}. Subsequent generalizations for lattices with dissipative interactions (e.g. plasticity and bond sliding) were provided in~\cite{BeexDisLatt,BeexFiber}. In principle, the QC is a numerical procedure that can deal with local lattice-level phenomena in small regions of interest, whereas the lattice model is coarse grained in the remainder of the domain.

\begin{figure}
	\centering
	\includegraphics[scale=1]{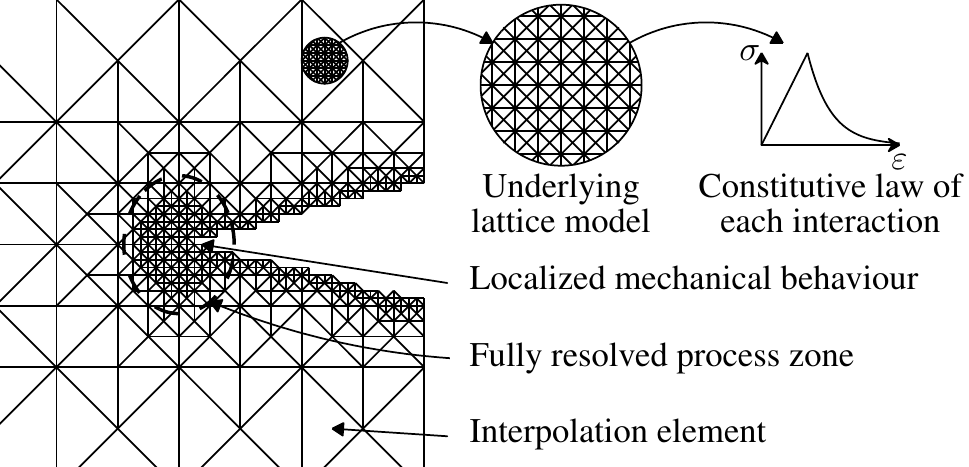}
	\caption{Sketch of a crack propagating through a lattice model using the variational adaptive QC method. At the crack tip and in the process zone, large fibre reorientations and deformations accompanied by dissipative processes may take place, which require the full lattice resolution. Elsewhere, the displacements of the underlying lattice can effectively be interpolated, which allows for coarse graining and numerical homogenization.}
	\label{Sect:Introduction:Fig:1}
\end{figure}

The aim of this contribution is to develop a QC framework that can deal with the initiation and subsequent propagation of damage and fracture in the underlying structural lattice model. Because such a phenomenon tends to be a highly localized and rather unstable process, sensitive to local mesh details, the QC framework must fully refine in critical regions \emph{before} any damage occurs in order to capture the physics properly, cf. Fig.~\ref{Sect:Introduction:Fig:1}. Moreover, as the entire problem is evolutionary, the location of the fully resolved region must evolve as well, which requires an adaptive QC framework. Several previous studies have also focused on adaptivity in QC methodologies, but they were always dealing with atomistic lattice models at the nano-scale, see e.g.~\cite{MillTad2002,Shenoy:1999,MemarnahavandiGoal}. Contrary to that, this contribution focuses on structural lattice networks for materials with discreteness at the meso-scale.

The QC approach aimed for is schematically presented in Fig.~\ref{Sect:Introduction:Fig:1}. The macro-scale fracture emerges as individual interactions' failures at the lattice level. Their damage leads to strain-softening and hence, the fracture process zone remains spatially localized. Consequently, only the crack tip and the process zone have to be fully resolved. The displacement fluctuations elsewhere remain small, allowing for efficient interpolation and coarse graining. Due to the spatial propagation of the crack front through the system of interest, available QC formulations need to be generalized to involve dissipation induced by damage, and an adaptive meshing scheme that includes a suitable marking strategy needs to be developed.

The theoretical framework employed in this contribution is closely related to our variational QC formulation for hardening plasticity discussed in~\cite{VarQCDiss}. The present work can be viewed as an extension towards lattices with localized damage and with an adaptive refinement strategy. In principle, the overall procedure is based on the variational formulation by Mielke and Roub\'{i}\v{c}ek~\cite{MieRou:2015}, developed for rate-independent inelastic systems. This variational formulation employs at each time instant~$t_k$ an incremental potential energy~$\Pi^k$, that can be minimized with respect to the observable (kinematic) as well as the internal (history, dissipative) variables. Hence, this formulation is different from the one employed in the virtual-power-based QC framework of Beex \emph{et al.}~\cite{BeexDisLatt}, which is based on the virtual-power statement of the lattice model in combination with a Coleman-Noll procedure. The theoretical concepts of the variational formulation and its application to damaging lattice models are discussed in Section~\ref{Sect:VarForm}.

After the incremental energy is presented for the full lattice system, two reduction steps can be applied to it in analogy to the standard QC framework, see e.g.~\cite{TadmorModel,Iyer:QC:2011,Luskin:QC:2013,EidStuk}. In the first step, \emph{interpolation} constrains the displacements of all atoms\footnote{Note that throughout this work the term "atom" is used to refer to individual lattice nodes or particles, consistently with the original QC terminology developed for atomistic systems.} according to the displacements of a number of selected representative atoms, or \emph{repatoms} for short. This procedure reduces the number of degrees of freedom drastically. In the second step, only a small number of atoms is sampled to approximate the exact incremental energy~$\Pi^k$, its gradients, and Hessians, analogously to the numerical integration of FE technology. This step is referred to as \emph{summation}, and it entails also a significant reduction of the number of internal variables. Together, the two steps yield a reduced state variable~$\bs{q}_\mathrm{red}\in\mathscr{Q}_\mathrm{red}$ and an approximate incremental energy~$\widehat{\Pi}(\bs{q}_\mathrm{red})$. In Section~\ref{Sect:QC}, a more detailed discussion about QC techniques, adaptive modelling, marking strategy, and mesh refinement will be provided. Consequences of the proposed mesh refinement strategy will be discussed also from the energetic point of view.

The minimization of~$\Pi^k$ provides governing equations presented in Section~\ref{Sect:Sol}, where a suitable solution strategy is described. Incremental energy minimization procedures often employ some version of the Alternating Minimization~(AM) strategy, see e.g.~\cite{BourdinVar,burke_adaptive_2010,Hofacker2012,NegKne15}. The approach used here, however, minimizes the so-called \emph{reduced energy}, i.e. the energy potential~$\Pi^k$ with eliminated internal variables, cf.~\cite{Carstensen2002:nonconvex} or~\cite{MieRou:2015}. As a result, the overall solution process simplifies and is more efficient compared to the AM approach.

In Section~\ref{Sect:Examples}, the proposed theoretical developments are first applied to an L-shaped plate test. The force-displacement diagrams and crack paths predicted with the adaptive QC approach are compared to those predicted with full lattice computations. Further, the energy consistency during the entire evolution is assessed and the errors in energies are discussed. The second numerical example focuses on the antisymmetric four-point bending test, described e.g. in~\cite{SchlangenThesis}. It demonstrates the ability of the adaptive QC scheme to predict nontrivial, curved crack paths. Finally, this contribution closes with a summary and conclusions in Section~\ref{Sect:Conclusion}.
%
%
\section{Variational Formulation of Lattice Structures with Damage}
\label{Sect:VarForm}
In this section, we recall the general variational theory for rate-independent systems, discussed e.g. in~\cite{Francfort1993stable,HanReddy,Francfort:1998,Ortiz:CMAME:1999,Charlotte}, followed by the geometric setting, description of state variables, and by the construction of energies. The entire exposition will be confined to 2D systems, but the extension to 3D is straightforward.
%
%
\subsection{General Considerations}
\label{SubSect:GenCon}
The evolution of a system within a time horizon~$[0,T]$ is considered to be \emph{quasistatic} and \emph{rate-independent}, so the (pseudo-) time~$t\in[0,T]$ can be arbitrarily rescaled without any influence on the results. The system of interest is fully specified by the potential (Gibbs type) energy~$\mathcal{E} : [0,T]\times\mathscr{Q}\rightarrow\mathbb{R}$ and by the dissipation distance~$\mathcal{D}(\bs{z}_2,\bs{z}_1)$, $\mathcal{D}:\mathscr{Z}\times\mathscr{Z}\rightarrow\mathbb{R}^+\cup\{+\infty\}$. The dissipation distance reflects the minimum energy dissipated by the continuous transition between two consecutive states~$\bs{z}_1$ and~$\bs{z}_2$. Both functionals are defined for states~$\bs{q}(t)=(\bs{r}(t),\bs{z}(t))\in\mathscr{Q}$, where~$\mathscr{Q}=\mathscr{R}\times\mathscr{Z}$ is a suitable state-space, $\bs{r}(t)\in\mathscr{R}$ is a set of observable kinematic variables, and~$\bs{z}(t)\in\mathscr{Z}$ is a set of internal variables describing the inelastic processes. 

A function~$\bs{q} : [0,T] \rightarrow \mathscr{Q}$ is called an \emph{energetic solution} of the energetic rate-independent system~$(\mathcal{E},\mathcal{D},\bs{q}_0)$ if it satisfies the following \emph{stability condition}~\eqref{S} and \emph{energy balance}~\eqref{E} for all~$t \in [0,T]$
\begin{align}
	&\mathcal{E}(t,\bs{q}(t))\leq\mathcal{E}(t,\widehat{\bs{q}})+\mathcal{D}(\widehat{\bs{z}},\bs{z}(t)), \quad \forall \widehat{\bs{q}} \in \mathscr{Q},\label{S}\tag{S}\\
	&\mathcal{V}(\bs{q}(t))+\mathrm{Var}_{\mathcal{D}}(\bs{q};0,t)=\mathcal{V}(\bs{q}(0))+\mathcal{W}_\mathrm{ext}(\bs{q};0,t),
	\tag{E}\label{E}
\end{align}
along with the \emph{initial condition}
\begin{equation}
\bs{q}(0)=\bs{q}_0.
\tag{I}
\label{I}
\end{equation}
For further details see e.g~\cite{MieRou:2015}. The potential energy~$\mathcal{E}$ can be expressed as
\begin{equation}
\mathcal{E}(t,\widehat{\bs{q}}) = \mathcal{V}(\widehat{\bs{q}}) - \left<\bs{f}(t),\widehat{\bs{q}}\right>,
\label{Sect:VarForm:Eq:0}
\end{equation}
where~$\mathcal{V} : \mathscr{Q}\rightarrow\mathbb{R}$ is the internal free (Helmholtz type) energy, $\bs{f} : [0,T] \rightarrow \mathscr{R}^* \times \mathscr{Z}^*$ represents an external loading (which satisfies some additional conditions related to the boundedness of~$\mathcal{E}$ not specified here, for further details see~\cite{MieRou:2015}), $\mathscr{R}^*$ and~$\mathscr{Z}^*$ are spaces dual to~$\mathscr{R}$ and~$\mathscr{Z}$, and~$\left<\bullet, \bullet\right>$ denotes the corresponding duality pairing. In~\eqref{E}, we have introduced the work performed by the external forces
\begin{equation}
\mathcal{W}_\mathrm{ext}(\bs{q};0,t) = \left< \bs{f}(t), \bs{q}(t) \right> - \left< \bs{f}(0), \bs{q}(0) \right> + \int_0^t \frac{\partial}{\partial s}\mathcal{E}(s,\bs{q}(s)) \,\mbox{d}s,\footnote{Note that in the case of externally prescribed forces, we have~$\frac{\partial}{\partial s}\mathcal{E}(s,\bs{q}(s)) = -\left<\dot{\bs{f}}(s),\bs{q}(s)\right>$, where the dot stands for the time derivative. Furthermore, assuming sufficient smoothness of all data, one obtains~$\mathcal{W}_\mathrm{ext}(\bs{q};0,t) = \int_0^t\left<\bs{f}(s),\dot{\bs{q}}(s)\right>\,\mathrm{d}s$ from integration by parts.}
\label{Sect:VarForm:Eq:1a}
\end{equation}
and the dissipated energy
\begin{equation}
\mathrm{Var}_{\mathcal{D}}(\bs{q};0,t) =  \sup\left\{\sum_{k=1}^{n}\mathcal{D}(\bs{z}(t_{k}),\bs{z}(t_{k-1}))\right\},
\label{Sect:VarForm:Eq:1b}
\end{equation}
where the symbol~$(\bs{q};0,t)$ indicates the dependence on~$\bs{q}(s)$ for~$s\in[0,t]$, and where the supremum is taken over all~$n\in\mathbb{N}$ and all partitions of the time interval~$[0,t]$.

Upon introducing a discretization of the time interval~$[0,T]$ in the form~$0=t_0<t_1<\dots<t_{n_T} = T$, the time-discrete energetic solution can be constructed by an \emph{incremental problem}~\eqref{IP} defined as
\begin{equation}
\bs{q}(t_k) \in \underset{\widehat{\bs{q}}\in\mathscr{Q}}{\mbox{arg min }}
\Pi^k(\widehat{\bs{q}};\bs{q}(t_{k-1})), \quad k = 1, \ldots, n_T,
\tag{IP}
\label{IP}
\end{equation}
In~\eqref{IP}, each step is realized as a minimization problem of the following \emph{incremental energy}:
\begin{equation}
\Pi^k(\widehat{\bs{q}};\bs{q}(t_{k-1}))=\mathcal{E}(t_k,\widehat{\bs{q}})+\mathcal{D}(\widehat{\bs{z}},\bs{z}(t_{k-1})).
\tag{IE}
\label{IE}
\end{equation}

A characteristic difficulty related to time-discrete energetic solutions is that they are constructed by a recursive global minimization, which is computationally cumbersome for non-convex energies. Yet it is reasonable and standard to assume that solutions of~\eqref{IP} are associated with local minima that satisfy the energy balance~\eqref{E}, cf. e.g.~\cite{Francfort:Fracture}. Such a strategy, in combination with indirect displacement solution control, is adopted also in this work. Local minimization entails that any additional requirements for~$\bs{f}$ can be dropped, cf. Eq.~\eqref{Sect:VarForm:Eq:0}. For further discussion about local versus global minimization see e.g.~\cite{Levitas}. Note that within the energetic framework, the solution~$\bs{q}$ may be a discontinuous function of time, leading to jumps in internal variables and energy quantities.
%
%
\subsection{Geometry, Kinematics, and Internal Variables}
\label{SubSect:Geometry}
\begin{figure}
	\centering
	\includegraphics[scale=1]{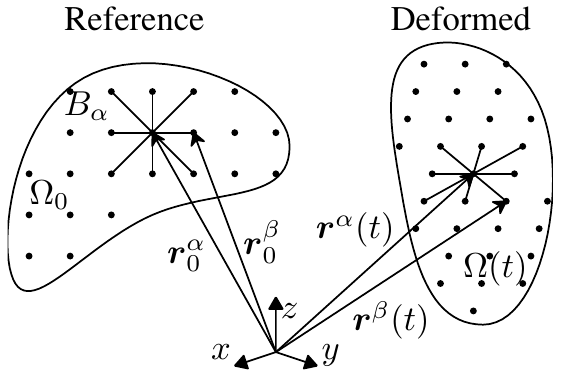}
	\caption{Sketch of geometric variables and two system configurations: reference configuration~$\Omega_0$ and current configuration~$\Omega(t)$.}
	\label{DissLatt:Fig:1}
\end{figure}

To specialize the abstract framework to lattice networks, we first introduce the geometric setting and notation, cf. Fig.~\ref{DissLatt:Fig:1}. The domain in a reference configuration~$\Omega_0=\Omega(0)\subset\mathbb{R}^{2}$ contains~$n_\mathrm{ato}$ atoms, collected in an index set~$N_\mathrm{ato}$. The spatial position of each atom~$\alpha \in N_\mathrm{ato}$ is specified by its position vector~$\bs{r}_0^\alpha\in\mathbb{R}^{2}$. Since only regular networks with nearest neighbour interactions are considered, the atoms' positions can be expressed as linear combinations of the primitive lattice vectors (in analogy to the Bravais lattices). All position vectors~$\bs{r}_0^\alpha$ are collected in a column matrix~$\bs{r}_0=[\bs{r}_0^1,\dots,\bs{r}_0^{n_\mathrm{ato}}]^\mathsf{T}$, $\bs{r}_0 \in \mathbb{R}^{2\, n_\mathrm{ato}}$. Throughout this contribution, Greek indices refer to atom numbers whereas Latin indices are reserved for spatial coordinates and other integer parametrizations. The nearest neighbours of an atom~$\alpha$ are furthermore stored in a set~$B_\alpha \subset N_\mathrm{ato}$. In contrast to atomistic lattices, the nearest neighbours of each atom do not change in time. The initial distance between two neighbouring atoms~$\alpha$ and~$\beta$ and the set of all initial distances between neighbouring atoms within the network are defined as
\begin{subequations}
	\label{SubSect:DissLatt:Eq:1}
	\begin{align}
	r_0^{\alpha\beta}(\bs{r}_0)&=||\bs{r}_0^\beta-\bs{r}_0^\alpha||_2,\label{SubSect:DissLatt:Eq:1a}\\
	\{r_0^{\alpha\beta}(\bs{r}_0)\}&=\{r_0^{\alpha\beta}\,|\,\alpha=1,\dots,n_\mathrm{ato},\ \beta \in B_\alpha,\mbox{ duplicity removed}\},\label{SubSect:DissLatt:Eq:1b}
	\end{align}
\end{subequations}
where~$||\bullet||_2$ is the Euclidean norm. Since~$r_0^{\alpha\beta}=r_0^{\beta\alpha}$, the set~$\{r_0^{\alpha\beta}\}$ in~\eqref{SubSect:DissLatt:Eq:1b} consists of~$n_\mathrm{int}$ components, where~$n_\mathrm{int}$ is the number of all interactions collected in an index set~$N_\mathrm{int}$. The above introduced symbol~$\alpha\beta$ will be employed below in two contexts. First, in the context of atoms, the symbol~$r_0^{\alpha\beta}$ measures the distance in the reference configuration between two atoms~$\alpha,\beta\in N_\mathrm{ato}$, as used in Eq.~\eqref{SubSect:DissLatt:Eq:1a}. Second, in the context of interactions, the same symbol~$r_0^{\alpha\beta}$ measures the length of the $p$-th interaction in the reference configuration, $p \equiv \alpha\beta$, $p \in N_\mathrm{int}$, with end atoms~$\alpha,\beta \in N_\mathrm{ato}$. A similar convention holds also for other physical quantities.

As the body deforms, the atoms in the reference configuration transform to a current configuration~$\Omega(t)\subset\mathbb{R}^2$. The deformed locations of all atoms are specified by their position vectors~$\bs{r}^\alpha(t)$, $\alpha=1,\dots,n_\mathrm{ato}$. In analogy to~$\bs{r}_0$, they are collected in a column matrix~$\bs{r}(t) = [\bs{r}^1(t),\dots,\bs{r}^{n_\mathrm{ato}}(t)]^\mathsf{T}$, $\bs{r}(t)\in\mathbb{R}^{2\, n_\mathrm{ato}}$, that represents also the abstract observable variable. The distance measure between two atoms~$r^{\alpha\beta}(\bs{r}(t))$ and the set of all distances~$\{r^{\alpha\beta}(\bs{r}(t))\}$ are introduced in the same manner as for the reference configuration, cf. Eq.~\eqref{SubSect:DissLatt:Eq:1b}. Due to kinematic boundary conditions, $\mathscr{R}(t)$ is a function of time and forms a manifold in~$\mathbb{R}^{2\, n_\mathrm{ato}}$.

Each interaction is endowed with one internal variable, a damage variable~$\omega^{\alpha\beta}(t)$, $0\leq\omega^{\alpha\beta}(t)\leq 1$. For brevity, all~$\omega^{\alpha\beta}(t)$ are collected in a column matrix~$\bs{z}(t)$, that represents the abstract dissipative variable. Recall that~$\bs{z} \in \mathscr{Z}$ meaning that~$\mathscr{Z}$ is identified with~$\mathbb{R}^{n_\mathrm{int}}$.
%
%
\subsection{Definition of Energies for Lattice Structures}
\label{SubSect:Energies}
The potential energy reads, cf. also Eq.~\eqref{Sect:VarForm:Eq:0},
\begin{equation}
\mathcal{E}(t,\widehat{\bs{q}})=\mathcal{V}(\widehat{\bs{r}},\widehat{\bs{z}})-\bs{f}_\mathrm{ext}^\mathsf{T}(t)\widehat{\bs{r}},
\label{SubSect:DissLatt:Eq:2}
\end{equation}
where the column matrix~$\bs{f}_\mathrm{ext}(t)\in\mathbb{R}^{2\, n_\mathrm{ato}}$ collects the prescribed external forces acting on lattice atoms. Note that in agreement with the notation used in Eqs.~\eqref{IP} and~\eqref{IE}, the hatted variables~$\widehat{\bullet}$ represent arbitrary admissible values, whereas non-hatted variables represent the minimizers of~\eqref{IP}. The internal free energy~$\mathcal{V}$ in~\eqref{SubSect:DissLatt:Eq:2} reflects the recoverable part of the energy stored in all interactions, and is expressed in terms of pair potentials
\begin{equation}
\mathcal{V}(\widehat{\bs{r}},\widehat{\bs{z}})=\widetilde{\mathcal{V}}(\{\widehat{r}^{\alpha\beta}(\widehat{\bs{r}})\},\widehat{\bs{z}})=\frac{1}{2}\sum_{\alpha,\beta\in B_\alpha} \left[ (1-\widehat{\omega}^{\alpha\beta})\phi^{\alpha\beta}(\widehat{r}^{\alpha\beta}_+) + \phi^{\alpha\beta}(\widehat{r}^{\alpha\beta}_-) \right],
\label{SubSect:DissLatt:Eq:3}
\end{equation}
where the first equality holds due to the principle of interatomic potential invariance, and where the factor~$1/2$ compensates for the fact that each pair of atoms appears twice in the sum over all~$\alpha$ and~$\beta$. Assuming~$\phi^{\alpha\beta}(r_0^{\alpha\beta}) = 0$, the two quantities~$\widehat{r}^{\alpha\beta}_+ = \max{(\widehat{r}^{\alpha\beta}, r^{\alpha\beta}_0)}$ and~$\widehat{r}^{\alpha\beta}_- = \min{(\widehat{r}^{\alpha\beta}, r^{\alpha\beta}_0)}$ in Eq.~\eqref{SubSect:DissLatt:Eq:3} ensure that interactions undergo damage only under tension and not in compression. For the conditions of damage growth, examined closely in Section~\ref{SubSect:FullSol}, this means that~$\omega^{\alpha\beta}$ can grow only if~$r^{\alpha\beta} > r^{\alpha\beta}_0$. The first term in the square brackets of Eq.~\eqref{SubSect:DissLatt:Eq:3}, i.e. the pair potential~$\phi^{\alpha\beta}$ weakened by~$(1-\omega^{\alpha\beta})$, reflects the elastic portion of the energy stored in a single interaction stretched to a length~$\widehat{r}^{\alpha\beta} > r^{\alpha\beta}_0$ and damaged to the level of~$\widehat{\omega}^{\alpha\beta}$. Note that for~$\widehat{\omega}^{\alpha\beta} = 0$, the interaction is completely intact (stores the full amount of energy), whereas for~$\widehat{\omega}^{\alpha\beta} = 1$ the interaction is fully damaged (no energy can be stored in it). The second term in the square brackets of Eq.~\eqref{SubSect:DissLatt:Eq:3} contributes only in compression, and the pair potential~$\phi^{\alpha\beta}$ represents the elastic part of the energy stored in a single interaction compressed to a length~$\widehat{r}^{\alpha\beta} < r^{\alpha\beta}_0$, independently of the level of damage~$\widehat{\omega}^{\alpha\beta}$.

The energy dissipated by a single interaction, $\mathcal{D}^{\alpha\beta}$, during the evolution between two consecutive states~$\widehat{\bs{z}}_1$ and~$\widehat{\bs{z}}_2$, is defined as
\begin{equation}
\mathcal{D}^{\alpha\beta}(\widehat{\bs{z}}_2,\widehat{\bs{z}}_1)=
\left\{
\begin{array}{ll}
{\displaystyle D^{\alpha\beta}(\widehat{\omega}_2^{\alpha\beta})-D^{\alpha\beta}(\widehat{\omega}_1^{\alpha\beta})} &\mbox{if}\ \widehat{\omega}_2^{\alpha\beta}\geq\widehat{\omega}_1^{\alpha\beta}\\
+\infty & \mbox{otherwise,}
\end{array}
\right.\quad\alpha\beta=1,\dots,n_\mathrm{int},
\label{SubSect:DissLatt:Eq:4}
\end{equation}
where~$D^{\alpha\beta}(\bullet)$ reflects the amount of energy dissipated during a unidirectional damage process up to a given state~$\bullet$. Consequently, $D^{\alpha\beta}$ must be increasing with~$D^{\alpha\beta}(0)=0$ and~$D^{\alpha\beta}(1) = g_{f,\infty}$, where $g_{f,\infty}$ represents the energy dissipated by the complete failure process. For a general derivation and further details see Sections~\ref{SubSect:FullSol}, \ref{Sect:Examples}, and~\ref{Sect:A}. In definition~\eqref{SubSect:DissLatt:Eq:4}, the value~$+\infty$ restricts the internal variables~$\omega^{\alpha\beta}$ to be only non-decreasing functions of time. The global dissipation distance then simply collects contributions of all interactions, i.e.
\begin{equation}
\mathcal{D}(\widehat{\bs{z}}_2,\widehat{\bs{z}}_1)=\frac{1}{2}\sum_{\alpha,\beta \in B_\alpha}\mathcal{D}^{\alpha\beta}(\widehat{\bs{z}}_2,\widehat{\bs{z}}_1),
\label{SubSect:DissLatt:Eq:6}
\end{equation}
where, in analogy to Eq.~\eqref{SubSect:DissLatt:Eq:3}, the factor~$1/2$ appears because each interaction is counted twice in the sum.

Based on the formulations of~$\mathcal{E}$ and~$\mathcal{D}$, it may be clear that the total incremental energy~$\Pi^k$ can be expressed in terms of the incremental energies of each interaction, $\widetilde{\pi}^k_{\alpha\beta}$, or in terms of the incremental site-energies of each atom, $\widehat{\pi}^k_\alpha$,
\begin{subequations}
	\label{SubSect:DissLatt:Eq:7}
	\begin{align}
	\widetilde{\pi}_{\alpha\beta}^k(\widehat{\bs{q}};\bs{q}(t_{k-1})) & = (1-\widehat{\omega}^{\alpha\beta})\phi^{\alpha\beta}(\widehat{r}^{\alpha\beta}_+) + \phi^{\alpha\beta}(\widehat{r}^{\alpha\beta}_-) + \mathcal{D}^{\alpha\beta}(\widehat{\omega}^{\alpha\beta},\omega^{\alpha\beta}(t_{k-1})), \quad {\alpha\beta}=1,\dots,n_\mathrm{int},\label{SubSect:DissLatt:Eq:7a}\\
	\widehat{\pi}_\alpha^k(\widehat{\bs{q}};\bs{q}(t_{k-1})) & =\frac{1}{2}\sum_{\beta \in B_\alpha}\widetilde{\pi}_{\alpha\beta}^k(\widehat{\bs{q}};\bs{q}(t_{k-1})), \quad \alpha=1,\dots,n_\mathrm{ato},\label{SubSect:DissLatt:Eq:7b}
	\end{align}
\end{subequations}
in the form
\begin{equation}
\Pi^k(\widehat{\bs{q}};\bs{q}(t_{k-1}))=\sum_{\alpha\beta=1}^{n_\mathrm{int}}\widetilde{\pi}_{\alpha\beta}^k(\widehat{\bs{q}};\bs{q}(t_{k-1}))-\bs{f}_\mathrm{ext}^\mathsf{T}(t_k)\widehat{\bs{r}}=\sum_{\alpha=1}^{n_\mathrm{ato}}\widehat{\pi}_\alpha^k(\widehat{\bs{q}};\bs{q}(t_{k-1}))-\bs{f}_\mathrm{ext}^\mathsf{T}(t_k)\widehat{\bs{r}}.
\label{SubSect:DissLatt:Eq:8}
\end{equation}
Both expressions in Eq.~\eqref{SubSect:DissLatt:Eq:8} will be used later on for the full-lattice, and in a slightly adjusted form also for the QC, computations. Specifically, for the minimization of the incremental energy~$\Pi^k$ (or its approximation~$\widehat{\Pi}^k$ in Eq.~\eqref{SubSect:Summ:Eq:2}) with respect to the kinematic variables~$\widehat{\bs{r}}$, the definition via the site-energies~\eqref{SubSect:DissLatt:Eq:7b} will be employed. For the minimization with respect to the internal variables~$\widehat{\bs{z}}$, the definition using the interaction energies~\eqref{SubSect:DissLatt:Eq:7a} will be used.
%
%
\section{Adaptive Quasicontinuum Methodology}
\label{Sect:QC}
This section extends the previously discussed theory to an adaptive QC scheme. First, the two standard QC steps---interpolation in Section~\ref{SubSect:Interpol} and summation in Section~\ref{SubSect:Summ}---are applied to the incremental energy~$\Pi^k$ at a fixed time step~$t_k$. Subsequently, a heuristic marking strategy and a mesh refinement algorithm are presented in Sections~\ref{SubSect:Adaptivity} and~\ref{SubSect:Refinement}. Finally, adaptivity is discussed from an energetic point of view in Section~\ref{SubSect:MeshEnergy}.
%
%
\subsection{Interpolation}
\label{SubSect:Interpol}
According to the standard QC theory, we introduce~$n_\mathrm{rep}$ \emph{repatoms} stored in an index set~$N_\mathrm{rep}^\mathrm{ato}\subseteq N_\mathrm{ato}$, that behave in analogy to FE nodes. The kinematic variables of the remaining atoms, $N_\mathrm{ato} \backslash N_\mathrm{rep}^\mathrm{ato}$, are interpolated using finite element shape functions constructed between repatoms. This can be expressed as
\begin{equation}
\widehat{\bs{r}}=\bs{\Phi}\widehat{\bs{r}}_\mathrm{rep},
\label{SubSect:Interpol:Eq:1}
\end{equation}
where~$\widehat{\bs{r}}_\mathrm{rep}\in\mathscr{R}_\mathrm{rep}(t)$ represents the column matrix with the positions of all repatoms in an arbitrary admissible configuration, and where the interpolation matrix~$\boldsymbol{\Phi} \in \mathbb{R}^{2\, n_\mathrm{ato} \times 2\, n_\mathrm{rep}}$ stores the basis vectors spanning~$\mathscr{R}_\mathrm{rep}(t)$ column-wise. Because~$\mathscr{R}_\mathrm{rep}(t)$ is a linear subspace of~$\mathscr{R}(t)$, the basis vectors are column matrices of length~$2\,n_\mathrm{ato}$ as elements of~$\mathscr{R}(t)$. Substitution of Eq.~\eqref{SubSect:Interpol:Eq:1} in Eq.~\eqref{IE} entails that the incremental energy becomes a function of~$\widehat{\bs{r}}_\mathrm{rep}$ and~$\bs{r}_\mathrm{rep}(t_{k-1})$, i.e.
\begin{equation}
\Pi^k(\widehat{\bs{r}},\widehat{\bs{z}};\bs{r}(t_{k-1}),\bs{z}(t_{k-1}))=\Pi^k(\bs{\Phi}\widehat{\bs{r}}_\mathrm{rep},\widehat{\bs{z}};\bs{\Phi}\bs{r}_\mathrm{rep}(t_{k-1}),\bs{z}(t_{k-1})),
\label{SubSect:Interpol:Eq:2}
\end{equation}
which reduces the number of degrees of freedom associated with the kinematic variable, i.e. from~$2\,n_\mathrm{ato}$ to~$2\,n_\mathrm{rep}$. The minimization in~\eqref{IP} with respect to~$\widehat{\bs{r}}\in\mathscr{R}(t)$ then becomes a minimization over some subspace~$\mathscr{R}_\mathrm{rep}(t)$, which reduces the computational effort if~$n_\mathrm{rep} \ll n_\mathrm{ato}$. 

In order to specify~$\bs{\Phi}$ in more detail, one introduces the standard FE triangulation~$\mathcal{T}_0$ (with elements~$K \in \mathcal{T}_0$) of~$\Omega_0$ with piecewise affine shape functions inside triangles. In the region of interest, the triangulation fully recovers the underlying lattice, whereas it coarsens elsewhere. The individual components of interpolation matrix~$\bs{\Phi}$ then read
\begin{equation}
\Phi_{(2\alpha-1)(2j-1)} = \Phi_{(2\alpha)(2j)} =
\left\{
\begin{aligned}
& \varphi_{\beta_j}(\bs{r}_0^\alpha) && \mbox{for}\ \alpha \in N_\mathrm{ato},\ \beta_j \in N_\mathrm{rep}^\mathrm{ato},\ j=1,\dots,n_\mathrm{rep}\\
& 0 && \mbox{otherwise,}
\end{aligned}
\right.
\label{SubSect:Interpol:Eq:3}
\end{equation}
where~$\beta_j$ denotes the $j$-th element of the set~$N_\mathrm{rep}^\mathrm{ato}$ according to its ordering, and~$\varphi_{\beta}(\bs{r}_0^\alpha)$ represents the shape function associated with a repatom~$\beta$ that is evaluated in the undeformed position of an atom~$\alpha$. Note that higher-order shape functions can also be used, cf. e.g.~\cite{BeexHO,BeexBeams,YangMMM}.
%
%
\subsection{Summation}
\label{SubSect:Summ}
The summation step involves the selection of a limited number of site-energies to sample the contributions of all atoms. The expression for the total incremental energy can be written as
\begin{equation}
\Pi^k(\widehat{\bs{q}};\bs{q}(t_{k-1})) \approx \widehat{\Pi}^k(\widehat{\bs{q}};\bs{q}(t_{k-1})) = \sum_{\alpha \in S_\mathrm{ato}}w_\alpha\widehat{\pi}_\alpha^k(\widehat{\bs{q}};\bs{q}(t_{k-1})) -\bs{f}_\mathrm{ext}^\mathsf{T}(t_k)\widehat{\bs{r}},
\label{SubSect:Summ:Eq:1}
\end{equation}
where~$S_\mathrm{ato} \subseteq N_\mathrm{ato}$ denotes a set of~$n_\mathrm{sam}^\mathrm{ato}$ sampling atoms. These atoms are chosen carefully in order to estimate the energy of the entire lattice, in analogy to numerical integration in FE implementations. Sampling atom~$\alpha \in S_\mathrm{ato}$ represents the contributions of~$w_\alpha$ atoms, including its own ($w_\alpha \geq 1$). With respect to FE technology, $w_\alpha$ is equivalent to the weight of Gauss integration point~$\alpha$. Because~$n_\mathrm{sam}^\mathrm{ato} \ll n_\mathrm{ato}$, the computational effort associated with the assembly of the energy, gradients, and Hessians is substantially reduced due to summation. Explicit instructions on how to choose the sampling atoms and how to compute their weights~$w_\alpha$ can be found e.g. in~\cite{BeexQC,BeexCSumRule,Amelang:2015}. In Section~\ref{Sect:Examples}, the so-called \emph{central summation} rule is used, cf.~\cite{BeexCSumRule}, which considers only the atoms at the element vertices and one near the center of the triangle. The vertex atoms represent only themselves (the so-called discrete sampling atoms with~$w_\alpha = 1$), whereas the central one is taken to be representative of atoms inside the element and close to element boundaries (the so-called central sampling atom with~$w_\alpha \geq 1$). If no internal atom exists, all boundary atoms are considered as discrete sampling atoms.

Similarly to~$S_\mathrm{ato}$, we introduce a set of~$n_\mathrm{sam}^\mathrm{int}$ sampling interactions stored in an index set~$S_\mathrm{int} \subseteq N_\mathrm{int}$, defined as all interactions connected to all sampling atoms; in analogy to Eq.~\eqref{SubSect:DissLatt:Eq:1b}, any duplicity is removed. Consequently, the summation in Eq.~\eqref{SubSect:Summ:Eq:1} can be again expressed as a sum over all sampling interactions~$\alpha\beta \in S_\mathrm{int}$, i.e. in analogy to Eq.~\eqref{SubSect:DissLatt:Eq:8}, we can write
\begin{equation}
\widehat{\Pi}^k(\widehat{\bs{q}};\bs{q}(t_{k-1})) = \sum_{\alpha\beta \in S_\mathrm{int}}\overline{w}_{\alpha\beta}\widetilde{\pi}_{\alpha\beta}^k(\widehat{\bs{q}};\bs{q}(t_{k-1}))-\bs{f}_\mathrm{ext}^\mathsf{T}(t_k)\widehat{\bs{r}} = \sum_{\alpha \in S_\mathrm{ato}}w_\alpha\widehat{\pi}_\alpha^k(\widehat{\bs{q}};\bs{q}(t_{k-1}))-\bs{f}_\mathrm{ext}^\mathsf{T}(t_k)\widehat{\bs{r}},
\label{SubSect:Summ:Eq:2}
\end{equation}
where~$\overline{w}_{\alpha\beta}$ denote the weight factors corresponding to interactions rather than to atom sites. Since all internal variables associated with interactions in~$N_\mathrm{int} \backslash S_\mathrm{int}$ become irrelevant, a reduced dissipative internal variable (associated only with the sampling interactions) $\bs{z}_\mathrm{sam}(t)\in\mathscr{Z}_\mathrm{sam}$ can be introduced, where~$\mathscr{Z}_\mathrm{sam}$ is identified with~$\mathbb{R}^{n_\mathrm{sam}^\mathrm{int}}$. This, in combination with the interpolation step, gives rise to the reduced state variable~$\bs{q}_\mathrm{red}(t)=(\bs{r}_\mathrm{rep}(t),\bs{z}_\mathrm{sam}(t))\in\mathscr{Q}_\mathrm{red}(t)$ and to the reduced abstract state space~$\mathscr{Q}_\mathrm{red}(t)=\mathscr{R}_\mathrm{rep}(t)\times\mathscr{Z}_\mathrm{sam}$.
%
%
\subsection{Marking Strategy}
\label{SubSect:Adaptivity}
In Sections~\ref{SubSect:Interpol}~-- \ref{SubSect:Summ}, a fixed triangulation~$\mathcal{T}_0$ of the reference domain~$\Omega_0$ has been assumed, meaning that~$\mathscr{Q}_\mathrm{red}(t)$ was a function of time only due to the evolving kinematic boundary conditions. For the lattice of interest, however, the location of damage growth evolves during each computation. As the aim is to let the initiation of damage occur only in fully resolved regions, the triangulation has to evolve as well. For this purpose, the adaptive procedure summarized in Alg.~\ref{SubSect:Adaptivity:Alg:1} changes the dimensionality of~$\mathscr{Q}_\mathrm{red}$ at each time step~$t_k$ due to mesh refinement. Consequently, $\mathscr{Z}(t_k)$, $N_\mathrm{rep}^\mathrm{ato}(t_k)$, $\bs{\Phi}(t_k)$, $S_\mathrm{ato}(t_k)$, and~$S_\mathrm{int}(t_k)$ become functions of time too.
\begin{algorithm}
	\caption{An adaptive scheme for the incremental QC problem.}
	\label{SubSect:Adaptivity:Alg:1}
	\centering
	\vspace{-\topsep}
	\begin{enumerate}[1:]
		\item Initialize the system: apply initial condition~$\bs{q}_0$ and construct initial (coarse) mesh~$\mathcal{T}_0$ with required information, e.g.~$N_\mathrm{rep}^\mathrm{ato}(t_0)$, $\bs{\Phi}(t_0)$, $S_\mathrm{ato}(t_0)$, $S_\mathrm{int}(t_0)$.
		\item \textbf{for~$k=1,\dots,n_T$}
		\begin{enumerate}[(i):]
			\item Apply the boundary conditions at time~$t_k$, $\mathcal{T}_k=\mathcal{T}_{k-1}$, $N_\mathrm{rep}^\mathrm{ato}(t_k) = N_\mathrm{rep}^\mathrm{ato}(t_{k-1})$, $\bs{\Phi}(t_k)=\bs{\Phi}(t_{k-1})$, $S_\mathrm{ato}(t_k) =S_\mathrm{ato}(t_{k-1})$, $S_\mathrm{int}(t_k) = S_\mathrm{int}(t_{k-1})$, etc.
			\item Equilibrate the unbalanced system, i.e. solve for~$\bs{q}_\mathrm{red}(t_k)\in\mathscr{Q}_\mathrm{red}(t_k)$ in~\eqref{IP} using Eq.~\eqref{SubSect:Interpol:Eq:1} substituted into the approximate incremental energy~$\widehat{\Pi}^k(\widehat{\bs{q}}_\mathrm{red};\bs{q}_\mathrm{red}(t_{k-1}))$ defined in Eq.~\eqref{SubSect:Summ:Eq:2}.
			\item For each coarse element~$K\in\mathcal{T}_k$ evaluate its \emph{indicator} and decide for possible refinement, cf. condition~\eqref{SubSect:Adaptivity:Eq:2}; all elements~$K$ marked for refinement are collected in a set~$\mathcal{I}\subseteq\mathcal{T}_k$.
			\item \textbf{If}~$\mathcal{I}\neq\emptyset$, refine current mesh, update the system information~$\mathcal{T}_k$, $N_\mathrm{rep}^\mathrm{ato}(t_k)$, $\bs{\Phi}(t_k)$, $S_\mathrm{ato}(t_k)$, $S_\mathrm{int}(t_k)$, etc., and return to~(ii) since the refined system is unbalanced.\\ \textbf{Else if}~$\mathcal{I}=\emptyset$, the mesh has converged; proceed to~(v).
			\item Store relevant output data for $k$-th time step: $\bs{q}_\mathrm{red}(t_k)$, $\mathcal{T}_k$, $N_\mathrm{rep}^\mathrm{ato}(t_k)$, $\bs{\Phi}(t_k)$, $S_\mathrm{ato}(t_k)$, $S_\mathrm{int}(t_k)$, etc.
			\item Proceed to the next time step.
		\end{enumerate}
		\item \textbf{end}
	\end{enumerate}
	\vspace{-\topsep}
\end{algorithm}

Given a triangulation~$\mathcal{T}_k$ at a time step~$t_k$, a \emph{marking strategy} decides in stage~(iii) of Alg.~\ref{SubSect:Adaptivity:Alg:1} which triangles should be refined. This procedure consists of the evaluation of a \emph{mesh indicator} for each triangle that is not fully refined yet. As the damage phenomenon is sensitive to local mesh details, it is convenient to fully refine the triangulation in critical regions \emph{before} any damage occurs there. To this end, each coarse element~$K$ is endowed with a subset of sampling interactions~$S^K_\mathrm{int} \subset S_\mathrm{int}$. For each of these interactions that is moreover loaded in tension (i.e.~$r^{\alpha\beta} > r^{\alpha\beta}_0$, since damage is assumed to evolve only in tension---recall Section~\ref{SubSect:Energies}), the elastic energy is computed.

As the indicators have to be evaluated only for coarse triangles, the elastic energies usually reduce to pair potential evaluations, i.e. to~$\phi^{\alpha\beta}(r^{\alpha\beta}_+)$. Then, a triangle~$K$ is marked for refinement if at least one interaction from~$S_\mathrm{int}^K$ has its pair potential evaluation~$\phi^{\alpha\beta}(r^{\alpha\beta}_+)$ higher than a given threshold~$\theta\,\phi_\mathrm{max}^{\alpha\beta}$, where~$\theta\in(0,1)$ specifies a certain safety margin and~$\phi_\mathrm{max}^{\alpha\beta}$ denotes the stored elastic energy threshold in the bond~$\alpha\beta$ at which damage starts to evolve. The safety margin~$\theta$ serves to control the accuracy of the QC method and the mesh refinement algorithm. Furthermore, in order to make sure that no damage occurs prior to full refinement, the associated internal variables are verified: if~$\omega^{\alpha\beta} > 0$ for some interaction~$\alpha\beta \in S_\mathrm{int}^K$, this triangle is marked for refinement, irrespective of the energy condition.

The above-described procedure is formalized as follows. For a coarse triangle~$K \in \mathcal{T}_k$, the set~$S_\mathrm{int}^K$ is defined as all sampling interactions that are situated at least partly inside the element~$K$,\footnote{Note that each element~$K$ is considered as a closed set. Consequently, the interactions and atoms lying on an element's edges or vertices are contained in that element.} i.e. as the interactions that
\begin{itemize}
	\item are connected to the central sampling atom of~$K$, $\{\alpha \in S_\mathrm{ato}\,|\,w_\alpha \geq 1,\bs{r}_0^\alpha\in K \backslash \partial K\}$, if a central sampling atom exists,
	\item are connected to the discrete sampling atoms of~$K$, $\{\alpha \in S_\mathrm{ato}\,|\,w_\alpha=1,\bs{r}_0^\alpha\in \partial K\}$ i.e. vertex or edge atoms, and their intersections with given triangle~$K$ have strictly positive lengths.
\end{itemize}
The energy criterion is expressed as
\begin{equation}
	\phi^{\alpha\beta}(r^{\alpha\beta}_+) \geq \theta\,\phi_\mathrm{max}^{\alpha\beta}, \quad \alpha\beta \in S_\mathrm{int}^K, \theta \in (0,1),
	\label{SubSect:Adaptivity:Eq:1}
\end{equation}
and the triangle indicator is evaluated as
\begin{equation}
\begin{aligned}
&&\mbox{If condition~\eqref{SubSect:Adaptivity:Eq:1} or }\omega^{\alpha\beta}>0\mbox{ holds at least for one bond }\alpha\beta \in S_\mathrm{int}^K\\
&&\Longrightarrow\mbox{ mark~$K$ for refinement, i.e. add~$K$ to~$\mathcal{I}$.}
\end{aligned}	
\label{SubSect:Adaptivity:Eq:2}
\end{equation}

The proposed marking strategy is based on heuristic considerations and supposedly also performs well for diffuse phenomena such as hardening plasticity. In Section~\ref{Sect:Examples} we will show that this methodology performs well for the lattice of interest if the safety margin~$\theta$ is sufficiently small ($\theta \leq 0.5$). Note that in the case of quadratic energy potentials~$\phi^{\alpha\beta}$, as introduced later in Section~\ref{Sect:Examples}, the choices~$\theta = 0.5$ and~$0.25$ are equivalent to stress levels of~$70.7\,\%$ and~$50.0\,\%$ of the tensile strength~$E\varepsilon_0$; cf. also Eqs.~\eqref{Sect:Examples:Eq:1}~-- \eqref{Sect:Examples:Eq:2} and the discussion thereof. 

As an alternative to mesh indicators, one could use \emph{error estimators} such as the \emph{goal-oriented} error estimator presented e.g. in~\cite{OdenGoal,Prudhomme_2006Goal}, or~\cite{MemarnahavandiGoal}. This approach will not be pursued further in this work and is left as a possible future challenge.
%
%
\subsection{Mesh Refinement}
\label{SubSect:Refinement}
\begin{figure}
	\centering
	\subfloat[initial mesh~$\mathcal{T}_k$, LEPP($K_0$)]{\includegraphics[scale=1]{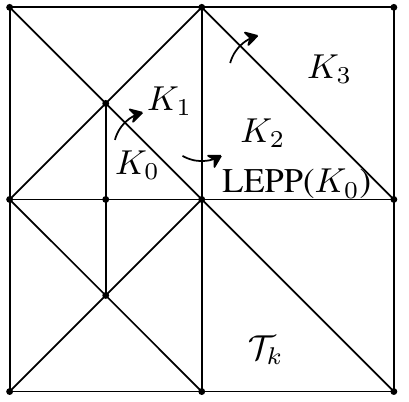}\label{SubSect:Adaptivity:Fig:1a}}
	\hspace{2em}
	\subfloat[final mesh]{\includegraphics[scale=1]{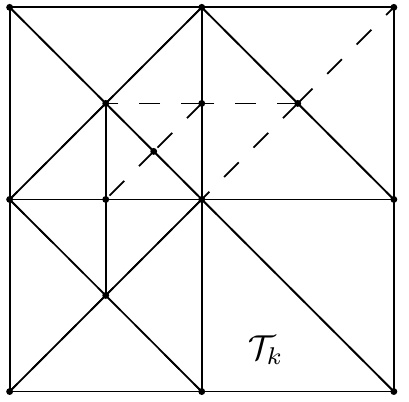}\label{SubSect:Adaptivity:Fig:1b}}
	\caption{Local refinement: (a)~initial mesh~$\mathcal{T}_k$ where triangle~$K_0$ is marked for refinement and its~LEPP($K_0$), (b)~final mesh with newly added edges in dashed lines.}
	\label{SubSect:Adaptivity:Fig:1}
\end{figure}
Elaborating on stage~(iv) of Alg.~\ref{SubSect:Adaptivity:Alg:1}, the current mesh~$\mathcal{T}_k$ is refined based on the marked triangles~$\mathcal{I}$. Since the underlying lattice is regular, it is convenient to employ a right-triangulated initial mesh~$\mathcal{T}_0$ and a self-similar mesh refinement. This may be considered reasonable because right-angled triangles lead to significantly smaller summation errors, cf.~\cite{VarQCDiss}, and because these triangles also naturally avoid artificial deformations in transition regions that would lead to spurious mesh refinements and non-physical evolution of internal variables. 

For the refinement, the standard Rivara~\cite{Rivara:LEPP} algorithm conserving \emph{non-degeneracy}, \emph{conformity}, and \emph{smoothness}, is used for each marked element ($K\in\mathcal{I}$). This algorithm tracks the so-called Longest-Edge Propagation Path associated with a triangle~$K$ in a backward manner, denoted for brevity as~LEPP($K$). For any given conforming triangulation~$\mathcal{T}_k$, the LEPP($K_0$) is defined as an ordered list of triangles~$K_0,K_1,\dots,K_n$, such that~$K_j$ is a neighbour to~$K_{j-1}$ by the longest edge of~$K_{j-1}$ for each~$j=1,\dots,n$. The Backward-Longest-Edge-Bisection algorithm for a pair~$(K_0,\mathcal{T}_k)$, recalled in Alg.~\ref{SubSect:Adaptivity:Alg:2}, is then used for each~$K \in \mathcal{I}$; for further details see e.g.~\cite{Rivara:LEPP}, Section~3. In Fig.~\ref{SubSect:Adaptivity:Fig:1}, the LEPP($K_0$) consists initially of four triangles, and the bisection proceeds from~$K_3$ towards~$K_0$.
\ignore{Several generalizations can be found in the literature. \cite{Suarez:RTIN} provided more progressive approach, the so-called four-triangle-longest-edge algorithm. Extensions to tetrahedra were presented by~\cite{RTIN:Tetrahedra}. Concerning implementation, the longest-edge-based refinement can be performed with the help of binary trees (see e.g.~\cite{Evans:RTIN}), quad trees (cf.~\cite{Pajarola:RTIN:QUAD} and~\cite{Pajarola:RTIN:QUAD1}), and with possible parallelization (cf.~\cite{RTIN:MPI}). These implementation techniques save the computational effort related to the transfer of data between meshes at different levels, updates in sampling atoms~$S_\mathrm{ato}$, internal variables~$\bs{z}$, interpolation matrix~$\bs{\Phi}$, etc.}
\begin{algorithm}
	\caption{Backward-Longest-Edge-Bisection algorithm for a pair~$(K_0,\mathcal{T}_k)$, $K_0\in\mathcal{I}\subseteq\mathcal{T}_k$.}
	\label{SubSect:Adaptivity:Alg:2}
	\centering
	\vspace{-\topsep}
	\begin{enumerate}[1:]
		\item Choose~$K_0\in\mathcal{I}$ scheduled for refinement.
		\item \textbf{while~$K_0$ is not bisected}
		\begin{enumerate}[(i):]
			\item Find/update the LEPP($K_0$).
			\item \textbf{If}~$K_n$, the last triangle of the LEPP($K_0$), is a terminal boundary triangle (its longest edge is a part of the physical boundary~$\partial\Omega_0$), bisect~$K_n$.\\
			\textbf{Else} bisect the last pair of terminal triangles of the LEPP($K_0$), $K_{n-1}$ and~$K_n$.
		\end{enumerate}
		\item \textbf{end}
	\end{enumerate}
	\vspace{-\topsep}
\end{algorithm}
%
%
\subsection{Energy Considerations}
\label{SubSect:MeshEnergy}
In this section, the implications of the adaptive scheme for the energy evolutions are discussed. The main motivations are threefold: (i)~the above presented theory is based on energy minimization, so the mesh refinement procedure should be consistent with these principles; (ii)~the energy balance~\eqref{E} must hold during the mesh refinement; (iii)~energy evolutions obtained with the adjusted mesh refinement compare much better to those computed for the full lattice solutions, hence the accuracy of the adaptive variational QC method can be assessed from the energetic viewpoint.

\begin{figure}
	\centering
	\includegraphics[scale=1]{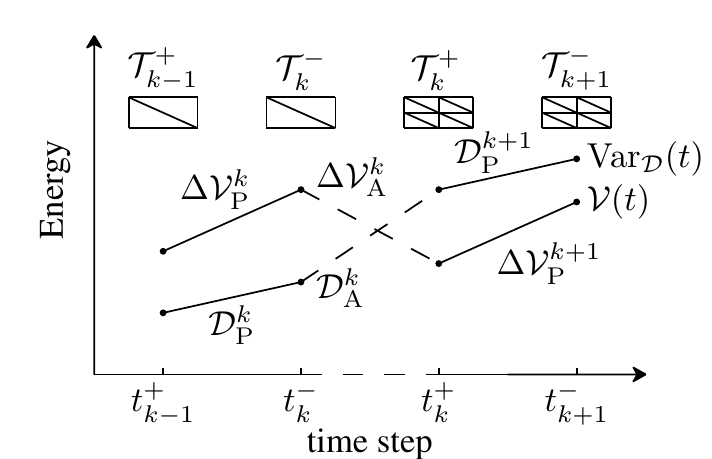}
	\caption{A sketch of the energy evolutions during mesh refinement at time step~$k$. For clarity, only the~$\mathcal{V}(t)$ and~$\mathrm{Var}_\mathcal{D}(t)$ energy components are shown.}
	\label{SubSect:MeshEnergy:Fig:1}
\end{figure}

First, the reader is referred to Fig.~\ref{SubSect:MeshEnergy:Fig:1}. Starting from a relaxed configuration at a time step~$t_{k-1}^+$ (for a converged mesh and an equilibrated system), the next load increment is applied (using the same triangulation of the previous time step, i.e.~$\mathcal{T}_k^- = \mathcal{T}_{k-1}^+$). This yields \emph{physical} energy increments, which are denoted by the subscript~"P", cf. Fig.~\ref{SubSect:MeshEnergy:Fig:1}. After the system is again equilibrated at~$t_k^-$ (using the same triangulation), the indicator condition~\eqref{SubSect:Adaptivity:Eq:2} is violated (assuming that the damage evolves) for some elements and hence, the mesh needs to be refined in several steps until convergence is reached at~$t_k^+$. Consequently, a new triangulation~$\mathcal{T}_k^+$ is obtained (stages (ii)~-- (iv) in Alg.~\ref{SubSect:Adaptivity:Alg:1}). During this mesh refinement (during the transition from~$t_k^-$ to~$t_k^+$), some elastic energy is released, which is referred to as \emph{artificial energy of constraints}; associated changes in energies are denoted by the subscript~"A", see Fig.~\ref{SubSect:MeshEnergy:Fig:1}. The elastic energy released due to mesh refinement was before (i.e. at~$t_k^-$) used to enforce some of the geometric constraints due to interpolation. The gradients of the released elastic energy (with respect to the kinematic variables) represent artificial constraining forces. The projection of these and the physical forces via~$\bs{\Phi}$ at~$t_k^-$ was zero (the system was in equilibrium). When the new triangulation~$\mathcal{T}^{+}_k$ is constructed, the system must be re-equilibrated because some of the geometric constraints are released. This takes place from stage~(iv) $\rightarrow$~(ii) in Alg.~\ref{SubSect:Adaptivity:Alg:1}. The elastic part of the artificial energy released during mesh refinement is denoted~$\Delta\mathcal{V}_\mathrm{A}^k$. This energy is transformed in three contributions: (i)~an additional dissipation increment~$\mathcal{D}_\mathrm{A}^k$, (ii)~a change of the work performed by the external reactions, and (iii)~work done due to the relaxation of the internal constraining forces. The latter two contributions are jointly denoted as~$\Delta\mathcal{W}_{\mathrm{ext},\mathrm{A}}^k$. 

Taking the above-described aspects into account and evaluating the internal elastic energy~$\mathcal{V}$, dissipated energy~$\mathrm{Var}_\mathcal{D}$, and the work done by the external forces~$\mathcal{W}_\mathrm{ext}$ at time instants~$t_k^+$ provides energy evolutions that we will call \emph{reconstructed}. Note that whereas the reconstructed QC energies can be compared well with the energy evolutions obtained for the fully-resolved system, the physical QC energies (energies evaluated at~$t_k^-$) are incomparable. This can be verified in Section~\ref{Sect:Examples} (Figs.~\ref{SubSect:SimpleEx:Fig:2b} and~\ref{SubSect:ComplexEx:Fig:4b}).

Finally, let us emphasize that the dissipation distance requires two states~$\widehat{\bs{z}}_1$ and~$\widehat{\bs{z}}_2$ (see definition~\eqref{SubSect:DissLatt:Eq:4}), that are inherently incompatible for~$t_k^-$ and~$t_k^+$ since they are associated with different triangulations~$\mathcal{T}_k^-$ and~$\mathcal{T}_k^+$. In order to compute~$\mathcal{D}_\mathrm{A}^k$, first the internal variables required for the new sampling interactions need to be established. In Section~\ref{Sect:Examples}, this is carried out such that the additional internal variables, $\omega^{\alpha\beta}$, $\alpha\beta \in S_\mathrm{int}(t_k^+) \backslash S_\mathrm{int}(t_k^-)$, are computed retrospectively by minimizing their incremental interaction energies, $\widetilde{\pi}^l_{\alpha\beta}$, $l < k$ including~$t_k^-$ (the entire evolution is required because~$\bs{z}(t)$ is history-dependent), with respect to internal variables while keeping the kinematic variables fixed ($\bs{r}(t_l)$ is computed from the previous results through Eq.~\eqref{SubSect:Interpol:Eq:1}). This operation corresponds to a mapping~$\bs{z}_\mathrm{sam} \rightarrow \bs{z}$ (not necessarily unique) touched upon in~\cite{VarQCDiss}, Section~3. Nonetheless, as the mesh is fully refined before any damage occurs (recall condition~\eqref{SubSect:Adaptivity:Eq:2}), all additionally required internal variables in all previous time steps are zero. An analogous procedure is applied also for the internal and external forces.
%
%
\section{Solution Strategy}
\label{Sect:Sol}
%
%
\subsection{Full-Lattice Computation}
\label{SubSect:FullSol}
Let us return momentarily from QC techniques to the full system, whose evolution is governed by~\eqref{IP} and discuss a suitable strategy for its solution. Because indirect displacement load control is used in Section~\ref{Sect:Examples}, it is convenient to adopt a minimization of the reduced incremental energy
\begin{equation}
\Pi^k_\mathrm{red}(\widehat{\bs{r}};\bs{q}(t_{k-1}))=\underset{\widehat{\bs{z}}\in\mathscr{Z}}{\mbox{min }}
\Pi^k(\widehat{\bs{r}},\widehat{\bs{z}};\bs{q}(t_{k-1})),
\label{Sect:Sol:Eq:1}
\end{equation}
see e.g.~\cite{MieRou:2015}, Section~3.1.2. In physical terms, the reduced energy~$\Pi^k_\mathrm{red}$ is obtained from the incremental energy~$\Pi^k$ in which certain atom positions~$\widehat{\bs{r}}$ are imposed. Computing for each interaction the damage that would occur at the strain that corresponds to the imposed atom positions and substituting the result back into~$\Pi^k$ provides~$\Pi^k_\mathrm{red}$. Taking into account Eq.~\eqref{Sect:Sol:Eq:1}, the incremental problem~\eqref{IP} transforms to
\begin{equation}
\bs{r}(t_k)=\underset{\widehat{\bs{r}}\in\mathscr{R}(t_k)}{\mbox{arg min }}
\Pi^k_\mathrm{red}(\widehat{\bs{r}};\bs{q}(t_{k-1}))
\tag{IP$_\mathrm{red}$}
\label{IPR}
\end{equation}
and can be solved by the standard Newton algorithm if~$\Pi_\mathrm{red}^k$ is sufficiently smooth in terms of~$\widehat{\bs{r}}$. The Taylor expansion of~$\Pi^k_\mathrm{red}$ in Eq.~\eqref{IPR} with respect to~$\widehat{\bs{r}}$ in the vicinity of~$\widehat{\bs{r}}^i$ provides the following stationarity condition
\begin{equation}
\bs{K}^i(\widehat{\bs{r}}^{i+1}-\widehat{\bs{r}}^i)+\bs{f}^i = \bs{0},
\label{SubSect:FullSol:Eq:1}
\end{equation}
where superscript~$i$ indicates the $i$-th iteration, and where
\begin{subequations}
\label{SubSect:FullSol:Eq:2}
\begin{align}
\bs{f}^i &= \bs{f}(\widehat{\bs{r}}^i) = \left.\frac{\partial\Pi_\mathrm{red}^k(\widehat{\bs{r}};\bs{q}(t_{k-1}))}{\partial\widehat{\bs{r}}}\right|_{\widehat{\bs{r}} = \widehat{\bs{r}}^i},\\
\bs{K}^i &= \bs{K}(\widehat{\bs{r}}^i) = \left.\frac{\partial^2\Pi_\mathrm{red}^k(\widehat{\bs{r}};\bs{q}(t_{k-1}))}{\partial\widehat{\bs{r}}\partial\widehat{\bs{r}}}\right|_{\widehat{\bs{r}} = \widehat{\bs{r}}^i}.
\end{align}
\end{subequations}
Eq.~\eqref{SubSect:FullSol:Eq:1} presents a system of linear equations for the increments~$\widehat{\bs{r}}^{i+1}-\widehat{\bs{r}}^i$. Iterating  Eqs.~\eqref{SubSect:FullSol:Eq:1} and~\eqref{SubSect:FullSol:Eq:2} until convergence of~$||\bs{f}^i||_2$ then yields~$\bs{r}(t_k)$. For completeness, we present explicit expressions for the gradients and Hessians in~\ref{Sect:B}.

For iteration~$i$ in Eq.~\eqref{SubSect:FullSol:Eq:1} the potential~$\Pi^k(\widehat{\bs{r}}^i,\widehat{\bs{z}};\bs{q}(t_{k-1}))$ needs to be minimized with respect to~$\widehat{\bs{z}}$ before the gradients and Hessians in Eq.~\eqref{SubSect:FullSol:Eq:2} are evaluated. This strategy is analogous to the condensation of internal variables in FE methods, because the Hessian ($\bs{K}^i$ in Eq.~\eqref{SubSect:FullSol:Eq:1}) effectively corresponds to the consistent tangent stiffness matrix. Problem~\eqref{Sect:Sol:Eq:1} for arbitrary fixed configuration~$\widehat{\bs{r}}$ is approached as follows: by rewriting the incremental energy into the interaction-wise form, cf. Eq.~\eqref{SubSect:DissLatt:Eq:8}, the minimization~\eqref{Sect:Sol:Eq:1} is split into separate one-dimensional problems that can be solved individually according to:
\begin{equation}
\mathring{\omega}^{\alpha\beta}=\underset{\omega^{\alpha\beta}(t_{k-1})\leq\widehat{\omega}^{\alpha\beta}\leq 1}{\mbox{arg min }}\widetilde{\pi}^k_{\alpha\beta}(\widehat{r}^{\alpha\beta},\widehat{\omega}^{\alpha\beta};\bs{q}(t_{k-1})), \quad \alpha\beta=1,\dots,n_\mathrm{int}.
\label{SubSect:FullSol:Eq:3}
\end{equation}
The box-constrained minimization~\eqref{SubSect:FullSol:Eq:3} coincides with the Karush--Kuhn--Tucker complementarity conditions for a uniformly stretched bar~$\alpha\beta$ with homogeneous damage. To see this, we take the derivative of~$\widetilde{\pi}^k_{\alpha\beta}$ in~\eqref{SubSect:FullSol:Eq:3} to obtain first-order optimality conditions:
\begin{equation}
\left(\left. \frac{\mathrm{d}}{\mathrm{d}\,\widehat{\omega}}\widetilde{\pi}^k_{\alpha\beta}(\widehat{r}^{\alpha\beta},\widehat{\omega};\bs{q}(t_{k-1}))\right|_{\widehat{\omega}=\mathring{\omega}^{\alpha\beta}}\right) \delta\omega \geq 0,
\quad \forall\,\delta\omega:\omega^{\alpha\beta}(t_{k-1})\leq\mathring{\omega}^{\alpha\beta}+\delta\omega\leq 1,
\label{SubSect:FullSol:Eq:4}
\end{equation}
see also~\cite[Section~3.1]{JiZe:Damage}. By recalling Eqs.~\eqref{SubSect:DissLatt:Eq:4} and~\eqref{SubSect:DissLatt:Eq:7a}, three states are distinguished:
\begin{enumerate}[(I)]
	\item $\omega(t_{k-1})=\mathring{\omega}^{\alpha\beta}<1$, i.e. elastic loading/unloading from a damaged state, implying~$0\leq\delta\omega\leq \underbrace{1-\mathring{\omega}^{\alpha\beta}}_{>0}$. Since~$\delta\omega$ is non-negative in this case, we obtain
	\begin{equation}
	\phi^{\alpha\beta}(\widehat{r}^{\alpha\beta}_+)\leq D'(\mathring{\omega}^{\alpha\beta}).
	\label{elastic}
	\end{equation}
	\item $\omega(t_{k-1})<\mathring{\omega}^{\alpha\beta}<1$, i.e. damage evolves, implying~$\underbrace{\omega(t_{k-1})-\mathring{\omega}^{\alpha\beta}}_{<0}\leq\delta\omega\leq\underbrace{1-\mathring{\omega}^{\alpha\beta}}_{>0}$. As~$\delta\omega$ can be either positive or negative, we have
	\begin{equation}
	\phi^{\alpha\beta}(\widehat{r}^{\alpha\beta}_+) = D'(\mathring{\omega}^{\alpha\beta}).
	\label{damaging}
	\end{equation}
	\item $\omega(t_{k-1})<\mathring{\omega}^{\alpha\beta}=1$, i.e. fully damaged state, implying~$\underbrace{\omega(t_{k-1})-\mathring{\omega}^{\alpha\beta}}_{<0}\leq\delta\omega\leq 0$. In this case, $\delta\omega$ can only be non-positive and hence,
	\begin{equation}
	\phi^{\alpha\beta}(\widehat{r}^{\alpha\beta}_+) \geq D'(\mathring{\omega}^{\alpha\beta}).
	\label{damaged}
	\end{equation}
\end{enumerate}
For brevity, we have denoted~$D'(\widehat{\omega})=\frac{\mathrm{d}}{\mathrm{d}\,\widehat{\omega}}\,D^{\alpha\beta}(\widehat{\omega})$. 

If the interaction undergoes compression, the minimization in Eq.~\eqref{SubSect:FullSol:Eq:3} is equivalent to
\begin{equation}
	\mathring{\omega}^{\alpha\beta} = \underset{\omega^{\alpha\beta}(t_{k-1})\leq\widehat{\omega}^{\alpha\beta}\leq 1}{\mbox{arg min }} D(\widehat{\omega}^{\alpha\beta}),
	\label{SubSect:FullSol:Eq:5}
\end{equation} 
which provides~$\mathring{\omega}^{\alpha\beta} = \omega(t_{k-1})$ as~$D$ is an increasing function of~$\widehat{\omega}^{\alpha\beta}$; recall definition in Eq.~\eqref{SubSect:DissLatt:Eq:4} and the discussion thereof. In accordance with the split of the internal energy~$\mathcal{V}$ into tensile and compressive parts specified in Eq.~\eqref{SubSect:DissLatt:Eq:3}, interaction~$\alpha\beta$ retains its full stiffness under compression, even for~$\mathring{\omega}^{\alpha\beta} > 0$.

Once the constitutive law is specified, i.e.~$D^{\alpha\beta}(\omega^{\alpha\beta})$ is given, internal variable~$\omega^{\alpha\beta}$ can be determined for stretched interaction~$\alpha\beta$ based on the three possible states~(I), (II), or~(III). The general derivation of~$D^{\alpha\beta}(\omega^{\alpha\beta})$ is presented in~\ref{Sect:A}, where its specific form for an exponential softening rule is derived as well; cf. also Section~\ref{Sect:Examples}.
%
%
\subsection{QC Computation}
\label{SubSect:QCSol}
Following the steps presented in Section~\ref{Sect:QC}, instead of directly minimizing the exact incremental energy~$\Pi_\mathrm{red}^k$, its approximation~$\Pi_\mathrm{red}^k\approx\widehat{\Pi}_\mathrm{red}^k$ is minimized with respect to the reduced variable~$\widehat{\bs{q}}_\mathrm{red}$. A fixed triangulation~$\mathcal{T}_k$ is assumed, meaning that only stage~(ii) of Alg.~\ref{SubSect:Adaptivity:Alg:1} is addressed in this section; the other ingredients of the algorithm have already been discussed and do not affect stage~(ii). Using the chain rule in the Taylor expansion of~$\widehat{\Pi}^k_\mathrm{red}$ provides the following stationarity condition
\begin{equation}
\bs{H}^i(\widehat{\bs{r}}^{i+1}_\mathrm{rep}-\widehat{\bs{r}}^i_\mathrm{rep})+\bs{G}^i = \bs{0},
\label{SubSect:QCSol:Eq:1}
\end{equation}
where
\begin{subequations}
\label{SubSect:QCSol:Eq:2}
\begin{align}
\bs{G}^i &= \bs{G}(\widehat{\bs{r}}^i_\mathrm{rep}) = \left.\bs{\Phi}^\mathsf{T}(t_k)\frac{\partial\widehat{\Pi}_\mathrm{red}^k(\widehat{\bs{r}};\bs{q}_\mathrm{red}(t_{k-1}))}{\partial\widehat{\bs{r}}}\right|_{\widehat{\bs{r}} = \bs{\Phi}(t_k)\widehat{\bs{r}}_\mathrm{rep}^i},\label{SubSect:QCSol:Eq:2a}\\
\bs{H}^i &= \bs{H}(\widehat{\bs{r}}^i_\mathrm{rep}) = \left.\bs{\Phi}^\mathsf{T}(t_k)\frac{\partial^2\widehat{\Pi}_\mathrm{red}^k(\widehat{\bs{r}};\bs{q}_\mathrm{red}(t_{k-1}))}{\partial\widehat{\bs{r}}\partial\widehat{\bs{r}}}\bs{\Phi}(t_k)\right|_{\widehat{\bs{r}} = \bs{\Phi}(t_k)\widehat{\bs{r}}_\mathrm{rep}^i},
\end{align}
\end{subequations}
and where the partial derivatives read
\begin{subequations}
\label{SubSect:QCSol:Eq:3}
\begin{align}
\frac{\partial\widehat{\Pi}_\mathrm{red}^k(\widehat{\bs{r}};\bs{q}_\mathrm{red}(t_{k-1}))}{\partial\widehat{\bs{r}}}&= -\bs{f}_\mathrm{ext}(t_k)+\sum_{\alpha \in S_\mathrm{ato}(t_k)}w_\alpha\frac{\partial\widehat{\pi}^k_{\mathrm{red},\alpha}(\widehat{\bs{r}};\bs{q}_\mathrm{red}(t_{k-1}))}{\partial\widehat{\bs{r}}}\label{SubSect:QCSol:Eq:3a}\\
&=-\bs{f}_\mathrm{ext}(t_k)+\sum_{\alpha \in S_\mathrm{ato}(t_k)}w_\alpha\bs{f}^\alpha_\mathrm{int}(\widehat{\bs{r}}),\nonumber\\
\frac{\partial^2\widehat{\Pi}_\mathrm{red}^k(\widehat{\bs{r}};\bs{q}_\mathrm{red}(t_{k-1}))}{\partial\widehat{\bs{r}}\partial\widehat{\bs{r}}}&=\sum_{\alpha \in S_\mathrm{ato}(t_k)}w_\alpha\frac{\partial^2\widehat{\pi}^k_{\mathrm{red},\alpha}(\widehat{\bs{r}};\bs{q}_\mathrm{red}(t_{k-1}))}{\partial\widehat{\bs{r}}\partial\widehat{\bs{r}}}=\sum_{\alpha \in S_\mathrm{ato}(t_k)}w_\alpha\bs{K}^\alpha(\widehat{\bs{r}}),
\end{align}
\end{subequations}
cf. also Eq.~\eqref{SubSect:Summ:Eq:1} and~\ref{Sect:A}. Note that in Eqs.~\eqref{SubSect:QCSol:Eq:3}, reduced site energies~$\widehat{\pi}^k_{\mathrm{red},\alpha}$ are introduced in analogy to~\eqref{Sect:Sol:Eq:1}. The converged solution of~\eqref{SubSect:QCSol:Eq:1} and~\eqref{SubSect:QCSol:Eq:2} yields~$\bs{r}_\mathrm{rep}(t_k)$. 

In order to construct~$\widehat{\Pi}^k_\mathrm{red}$, the interaction-wise formulation of the incremental energy is used, recall Eq.~\eqref{SubSect:Summ:Eq:2}. The minimization then requires the solution of~$n_\mathrm{sam}^\mathrm{int}(t_k)$ independent problems
\begin{equation}
\mathring{\omega}^{\alpha\beta}=\underset{\omega^{\alpha\beta}(t_{k-1})\leq\widehat{\omega}^{\alpha\beta}\leq 1}{\mbox{arg min }}\overline{w}_{\alpha\beta}\widetilde{\pi}^k_{\alpha\beta}(\widehat{r}^{\alpha\beta},\widehat{\omega}^{\alpha\beta};\bs{q}(t_{k-1})), \quad {\alpha\beta} \in S_\mathrm{int}(t_k).
\label{SubSect:QCSol:Eq:4}
\end{equation}
Because the solutions of~\eqref{SubSect:QCSol:Eq:4} do not depend on the weights~$\overline{w}_{\alpha\beta}$, each minimization problem can be solved independently, according to Section~\ref{SubSect:FullSol}.
%
%
\subsection{Boundary Conditions and Solution Control}
\label{SubSect:BCs}
Section~\ref{Sect:VarForm} made clear that the energetic solution~$\bs{q}(t)$ has to satisfy the energy equality~\eqref{E}, which serves as a selection criterion with respect to the numerous local minimizers of~$\Pi^k$. Usually, a certain variant of the \emph{backtracking} algorithm helps to select the proper solution, see e.g.~\cite{Bourdin:2007,MielkeZem,Benesova:2011,Mesgarnejad:2015}. In order to keep our exposition brief and simple, an alternative approach is followed. 

It can be shown that as long as the solution process remains continuous in time, the energy balance~\eqref{E} holds, see e.g.~\cite{StabPhaMarMaur,Pham:2013}. For this purpose, indirect load-displacement control is used in this contribution (see e.g.~\cite{Jirasek:2002}, Section~22.2.3), employing either Crack Mouth Opening Displacement (CMOD), Crack Mouth Sliding Dispacement (CMSD), or a combination of both. 

For the full lattice system, kinematic boundary conditions are imposed in the usual way, whereas prescribed forces enter the system via Eq.~\eqref{SubSect:DissLatt:Eq:2}, as the work performed by the force vector~$\bs{f}_\mathrm{ext}$. The indirect displacement control is then introduced via an additional parameter~$\lambda(t_k) \in \mathbb{R}$ that proportionally scales the prescribed displacements or forces. In case of prescribed forces for example, we have~$\bs{f}_\mathrm{ext}(t_k) = \lambda(t_k)\overline{\bs{f}}_\mathrm{ext}$, where~$\overline{\bs{f}}_\mathrm{ext}$ is a given reference load vector. The parameter~$\lambda(t_k)$ is then determined from the following scalar condition
\begin{equation}
(\bs{c}_\mathrm{o} + \bs{c}_\mathrm{s})^\mathsf{T}(\widehat{\bs{r}}^{i+1}-\bs{r}(t_{k-1})) = \overline{\Delta\ell},
\label{SubSect:SimpleEx:Eq2}
\end{equation}
which holds for prescribed step size~$\overline{\Delta\ell}$ at each Newton iteration and time step. Due to the substitution of~$\bs{f}_\mathrm{ext}(t_k) = \lambda(t_k)\overline{\bs{f}}_\mathrm{ext}$ in~\eqref{SubSect:DissLatt:Eq:2}, the solution increment in~\eqref{SubSect:FullSol:Eq:1} becomes a function of~$\lambda(t_k)$, which can subsequently be eliminated using Eq.~\eqref{SubSect:SimpleEx:Eq2}. A similar procedure is also applied to the kinematic boundary conditions. 

The column matrices~$\bs{c}_\mathrm{o}$ and~$\bs{c}_\mathrm{s}$ in Eq.~\eqref{SubSect:SimpleEx:Eq2} (both in~$\mathbb{R}^{2\, n_\mathrm{ato}}$) specify suitable displacement measures. The displacement difference between a pair of crack-mouth atoms in the direction perpendicular to a crack is specified by~$\bs{c}_\mathrm{o}$, i.e.~$\mathrm{CMOD} = \bs{c}_\mathrm{o}^\mathsf{T}(\widehat{\bs{r}}^{i+1}-\bs{r}(t_{k-1}))$, whereas~$\bs{c}_\mathrm{s}$ specifies the difference along the crack direction, i.e.~$\mathrm{CMSD} = \bs{c}_\mathrm{s}^\mathsf{T}(\widehat{\bs{r}}^{i+1}-\bs{r}(t_{k-1}))$. The combination of both simply reads~$\mathrm{CMOD}+\mathrm{CMSD} = (\bs{c}_\mathrm{o} + \bs{c}_\mathrm{s})^\mathsf{T}(\widehat{\bs{r}}^{i+1}-\bs{r}(t_{k-1}))$.

In the case of QC systems, the controlled quantity is~$\bs{r}_\mathrm{rep}(t_k)$, i.e. Eq.~\eqref{SubSect:SimpleEx:Eq2} is enforced for~$\bs{r}_\mathrm{rep}$, and the column matrices~$\bs{c}_\mathrm{o}$ and~$\bs{c}_\mathrm{s}$ are from~$\mathbb{R}^{2\,n_\mathrm{rep}}$. Finally, kinematic boundary conditions are applied only on repatoms and, in accordance with Eqs.~\eqref{SubSect:QCSol:Eq:1}, \eqref{SubSect:QCSol:Eq:2a}, and~\eqref{SubSect:QCSol:Eq:3a}, the QC system is loaded with~$\bs{\Phi}^\mathsf{T}(t_k)\bs{f}_\mathrm{ext}(t_k)$ if external forces are present.
\ignore{As an alternative to the indirect displacement control, dissipation- or energy-based versions of the arc-length method can be used, see~\cite{Verhoosel:2008} or~\cite{Vignollet:2015} for further details. One could possibly control displacement increments of interpolated atoms instead of repatoms; that would generally not be a good idea since interpolated displacements underestimate actual ones leading eventually to excessively large increments of~$\lambda$ and abrupt mesh refinements.}
%
%
\section{Numerical Examples}
\label{Sect:Examples}
The previously discussed theory is demonstrated on two benchmark problems in this section. A regular X-braced lattice with the following quadratic pair potential is used in both cases (the superscripts~$\alpha\beta$ are dropped for brevity)
\begin{equation}
\phi(\widehat{r}) = \frac{1}{2}\frac{EA}{r_0}\left(\widehat{r}-r_0\right)^2,
\label{Examples:Eq:1}
\end{equation}
where the interaction stiffness is~$EA/r_0$ according to standard truss theory, $E$ is the Young's modulus, and~$A$ the cross-sectional area. Note that this definition corresponds to the rotated engineering deformation measure and an undamaged bond. If damage takes place, $\phi$ needs to be multiplied by~$(1-\widehat{\omega})$, recall Eq.~\eqref{SubSect:DissLatt:Eq:3}.

\begin{figure}
	\centering
	\includegraphics[scale=1]{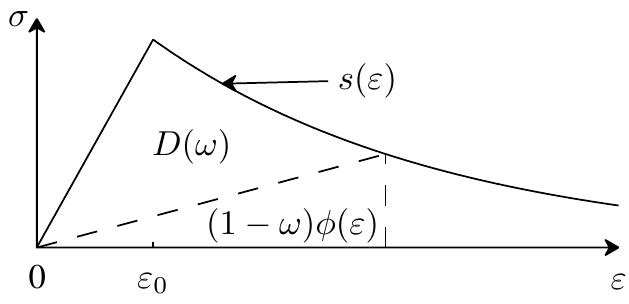}
	\caption{Exponential softening under tension: a sketch of the stress--strain diagram with corresponding quantities~$\varepsilon_0$ and~$s(\varepsilon)$, cf. Eq.~\eqref{Examples:Eq:2}. The elastically stored energy, $(1-\omega)\phi(\varepsilon)$, corresponds to the area of the dashed triangle, and the dissipated energy, $D(\omega)$, to the area of the upper triangle with the curved side.}
	\label{Sect:Examples:Fig:1}
\end{figure}

The mechanical behaviour is made independent of the sectional area and the initial length of the interaction by introducing bond strain~$\varepsilon$ and stress~$\sigma$ (see~\ref{Sect:A} for more details). In both examples below, the exponential stress-strain softening law sketched in Fig.~\ref{Sect:Examples:Fig:1} is employed. The strain softening branch~$s(\widehat{\varepsilon})$ takes the following form
\begin{equation}
s(\widehat{\varepsilon}) = E\varepsilon_0\exp{\left(-\frac{\widehat{\varepsilon}-\varepsilon_0}{\varepsilon_f}\right)}, \quad \varepsilon_0 \leq \widehat{\varepsilon},
\label{Examples:Eq:2}
\end{equation}
where~$\varepsilon_f$ is inversely proportional to its initial slope, and~$\varepsilon_0$ is the limit elastic strain. As shown in Eqs.~\eqref{pstress}~-- \eqref{SubSect:FullSol:Eq:9} of~\ref{Sect:A}, function~$s(\widehat{\varepsilon})$ in Eq.~\eqref{Examples:Eq:2} fully defines~$D(\widehat{\omega})$. The employed physical constants for both examples are specified in Tab.~\ref{Sect:Examples:Tab:1}.
\begin{table}
	\caption{Dimensionless material and geometric parameters for both test examples.}
	\centering
	\renewcommand{\arraystretch}{1.5}
	\begin{tabular}{l|r@{}lr@{}l}
		Physical parameters                                    & \multicolumn{2}{c}{Example~1} &                \multicolumn{2}{c}{Example~2}                \\ \hline
		Young's modulus, \hfill $E$                            & 1 &                           & 1 &  \\
		Cross-sectional area, \hfill $A$                       & 1 &                           & 1 &  \\
		Lattice spacing, \hfill along~$x$ and~$y$              & 1 &                           & 1 &  \\
		Limit elastic strain, \hfill $\varepsilon_0$           & 0 & .1                        & 0 & .01                                                     \\
		Inverse of initial slope, \hfill $\varepsilon_f$                  & 0 & .25                       & 0 & .025                                                    \\
		Displacement increment, \hfill $\overline{\Delta\ell}$ & 0 & .025                      & \multicolumn{2}{c}{see Fig.~\ref{SubSect:ComplexEx:Fig:2a}}
	\end{tabular}
	\label{Sect:Examples:Tab:1}
\end{table}
%
%
\subsection{L-Shaped Plate Example}
\label{SubSect:SimpleEx}
\begin{figure}
	\centering
	\includegraphics[scale=1]{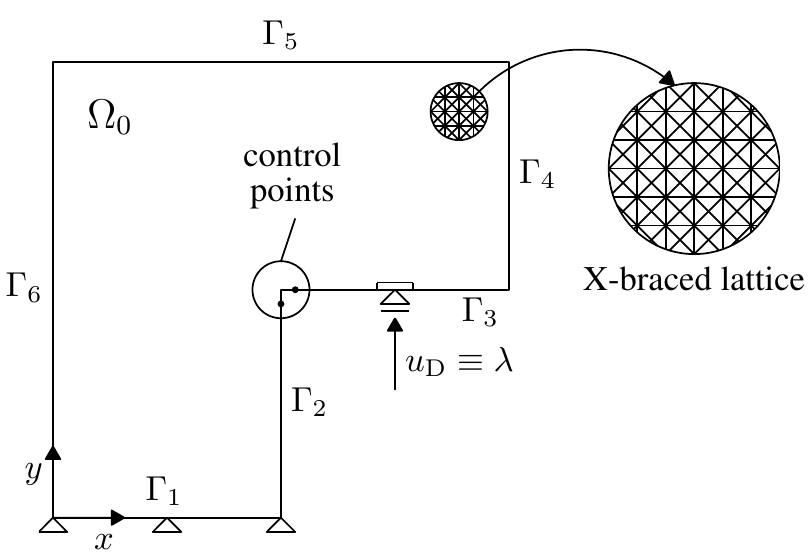}
	\caption{Sketch of the L-shaped plate example: geometry and boundary conditions. Variable~$u_\mathrm{D}$ denotes the applied vertical displacement.}
	\label{SubSect:SimpleEx:Fig:0}
\end{figure}
\begin{figure}
	\centering
	\subfloat[$\bs{r}(t)$ for~$u_\mathrm{D} = 14$]{\includegraphics[scale=1]{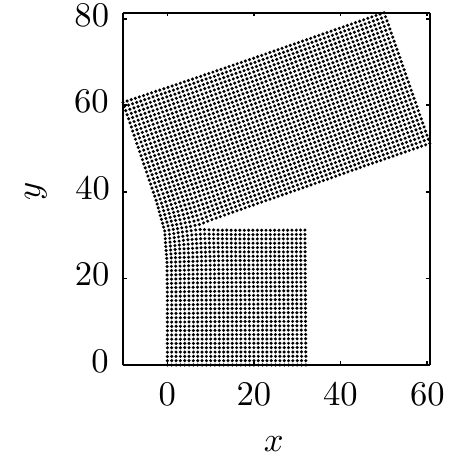}\label{SubSect:SimpleEx:Fig:1a}}\hspace{1em}
	\subfloat[force-displacement diagram]{\includegraphics[scale=1]{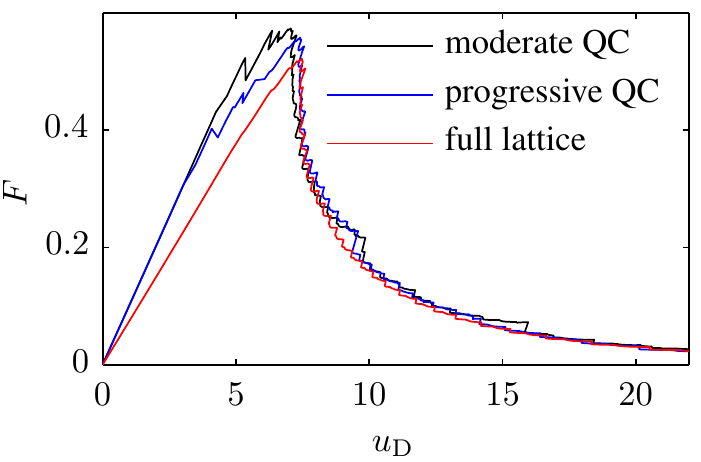}\label{SubSect:SimpleEx:Fig:1b}}	
	\caption{L-shaped plate test: (a)~deformed configuration for~$u_\mathrm{D} = 14$ (true, unscaled displacements are shown; note that only atoms are shown), and~(b) force-displacement diagram for reaction force~$F$ acting at~$\bs{r}(\Gamma_3/2)$ plotted against~$u_\mathrm{D}$.}
	\label{SubSect:SimpleEx:Fig:1}
\end{figure}

The first example considers an L-shaped plate modelled by a moderate number of atoms, serving therefore as a proof of concept. For a continuum-based analogy see e.g.~\cite{Mesgarnejad:2015}, Section~4.1. The reference domain~$\Omega_0$ fills a~$64\times 64$ lattice-spacing-sized square with cut out right-bottom quarter (see Fig.~\ref{SubSect:SimpleEx:Fig:0}). It comprises~$3,201$ atoms and~$12,416$ interactions. The lattice spacing is~$1$ unit length in both directions, and the material is homogeneous throughout the body. The specimen is fixed at the bottom part of its boundary, $\Gamma_1$, while a vertical displacement~$u_D$ is prescribed at~$\Gamma_3/2$, i.e.
\begin{subequations}
\label{SubSect:SimpleEx:Eq1}
\begin{align}
\bs{r}(\Gamma_1) &= \bs{r}_0(\Gamma_1),\\
r_y(\Gamma_3/2) &= r_{0,y}(\Gamma_3/2)+u_\mathrm{D} = 32+u_\mathrm{D}, \quad u_\mathrm{D} \geq 0,
\end{align}
\end{subequations}
where~$\partial\Omega_0=\bigcup_{i=1}^6\Gamma_i$, $\bs{r}(\Gamma)$ denotes the deformed positions of all atoms lying on a line segment~$\Gamma$, and~$\Gamma/2$ stands for the middle point of~$\Gamma$; each component of~$\bs{r}$ is a vector, i.e.~$\bs{r}^\alpha=[r_x^\alpha,r_y^\alpha]^\mathsf{T}$. In order to prevent any damage evolution due to the boundary condition being applied to a single point, local stiffening is used in the vicinity of~$\Gamma_3/2$, which is indicated by a small rectangle above the support in Fig.~\ref{SubSect:SimpleEx:Fig:0} (the Young's modulus is~$1000$ times larger than elsewhere and the limit elastic strain~$\varepsilon_0$ is infinite). The overall evolution of the system is controlled by the difference between the vertical positions of the two control points shown as black dots in Fig.~\ref{SubSect:SimpleEx:Fig:0}. After the crack initiates, cf. Fig.~\ref{SubSect:SimpleEx:Fig:1a}, this choice effectively corresponds to CMOD control; recall Eq.~\eqref{SubSect:SimpleEx:Eq2} with~$\bs{c}_\mathrm{s}=\bs{0}$.

The numerical study has been performed for three systems. The first one is the fully-resolved lattice, whereas the other two correspond to the adaptive QC with different safety margins~$\theta$, cf. Eq.~\eqref{SubSect:Adaptivity:Eq:1}. The first QC system uses a \emph{moderate} ($\theta = 0.5$) and the second a \emph{progressive} ($\theta = 0.25$) mesh refinement strategy; these systems are referred to as \emph{moderate QC} and \emph{progressive QC} for brevity. Rather low values of~$\theta$ are necessary because of the steep peaks of interaction (and site) energies occurring in the vicinity of the crack tip.

The deformed configuration predicted by the full-lattice computation at~$u_\mathrm{D} = 14$ is depicted in Fig.~\ref{SubSect:SimpleEx:Fig:1a}; note that only the atoms are shown. It can be observed that the crack initiates at the inner corner ($\Gamma_2 \cap \Gamma_3$), and propagates horizontally leftward. For all three systems, the crack paths are identical (not shown). The reaction force~$F$ as a function of~$u_\mathrm{D}$ is plotted in Fig.~\ref{SubSect:SimpleEx:Fig:1b}. In the initial stages, the reaction forces increase linearly, with the smallest stiffness corresponding to the fully-resolved system. Because both QC systems have the same initial triangulations~$\mathcal{T}_0$ (cf. Figs.~\ref{SubSect:SimpleEx:Fig:3a} and~\ref{SubSect:SimpleEx:Fig:3e}), their stiffnesses are identical. For an increasing applied displacement~$u_\mathrm{D}$, the progressive QC starts to refine first, followed by the moderate one. As the refinement process entails the relaxation of the geometric interpolation constraints and hence, a decrease of stiffness, the reaction forces approach the values of the fully-resolved system. The peak force is nevertheless overestimated by approximately~$10\,\%$ by both QC systems. The post-peak behaviour exhibits a mild structural snap-back, and the curves match satisfactorily with negligible error for the post-peak part of the diagram (when the crack fully localizes). Overall, it can be concluded that the results are more accurate for a lower safety margin~$\theta$. This entails, however, an increase in the number of repatoms and hence, decrease in efficiency. Nevertheless, once the crack fully localizes, the differences between the results for various values of~$\theta$ decrease, though the lower value is generally more accurate.
\begin{figure}
	\centering
	\subfloat[energy evolutions]{\includegraphics[scale=1]{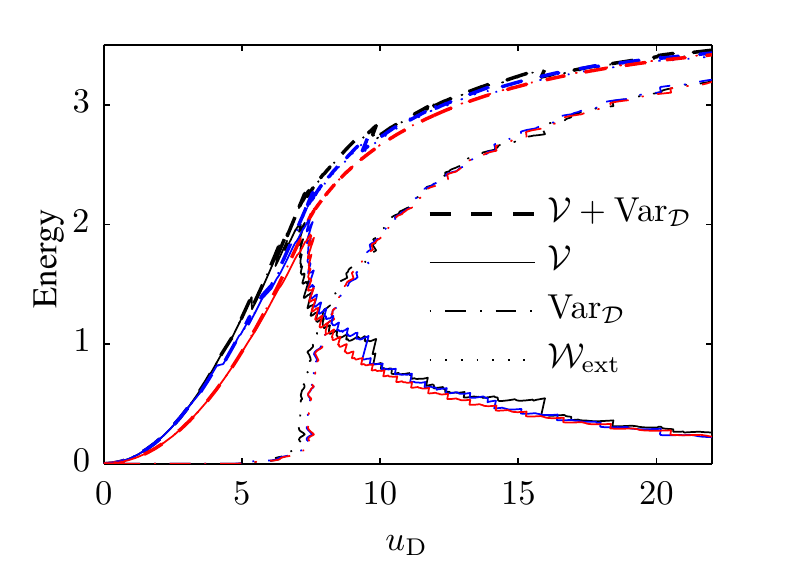}\label{SubSect:SimpleEx:Fig:2a}}\hfill
	\subfloat[energy components for moderate approach]{\includegraphics[scale=1]{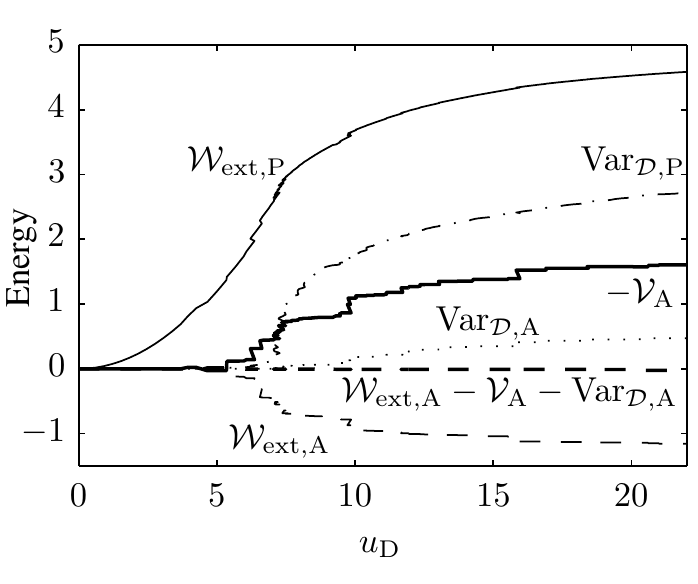}\label{SubSect:SimpleEx:Fig:2b}}
	\caption{Energy evolutions for the L-shaped plate test. (a)~Reconstructed energy evolution paths (black~-- the moderate QC approach~$\theta = 0.5$, blue~-- the progressive QC approach~$\theta = 0.25$, red~-- full-lattice solution). (b)~Energies exchanged during mesh refinement for the moderate QC approach, see Section~\ref{SubSect:MeshEnergy}.}
	\label{SubSect:SimpleEx:Fig:2}
\end{figure}

The energy evolution paths are shown in Fig.~\ref{SubSect:SimpleEx:Fig:2a} as functions of~$u_\mathrm{D}$. The elastic energy~$\mathcal{V}$ increases quadratically until the peak load and then drops gradually as the fracture process occurs. Near the peak load, the dissipated energy~$\mathrm{Var}_\mathcal{D}$ increases rapidly. It then continues to grow at a more moderate rate throughout the softening part of the load-displacement response. Notice that the energy balance~\eqref{E} is satisfied along the entire loading path, since the thin dotted lines corresponding to the work performed by external forces~$\mathcal{W}_\mathrm{ext}$ lie on top of the thick dashed lines representing~$\mathcal{V} + \mathrm{Var}_\mathcal{D}$. The maximum relative unbalance between~$\mathcal{V}+\mathrm{Var}_\mathcal{D}$ and~$\mathcal{W}_\mathrm{ext}$ is below~$1\,\%$. From the energy point of view, it can be concluded that both QC systems approximate the results of the full lattice system well. From the $\mathrm{Var}_\mathcal{D}$-curves we deduce that the crack starts to propagate first according to the moderate QC approach (because of its highest stiffness), and that the cracks corresponding to the progressive QC and to the full-lattice solution initiate almost at the same instant.

Let us recall Section~\ref{SubSect:MeshEnergy} and note that the energy evolutions presented in Fig.~\ref{SubSect:SimpleEx:Fig:2a} correspond to reconstructed energies evaluated at time instants~$t_k^+$, cf. Fig.~\ref{SubSect:MeshEnergy:Fig:1}. The energy components that are exchanged (the artificial energies) during the moderate QC prediction are presented in Fig.~\ref{SubSect:SimpleEx:Fig:2b}. We see that their magnitudes are large compared to the two physical energies ($\mathcal{W}_\mathrm{ext,P}$ and~$\mathrm{Var}_{\mathcal{D},\mathrm{P}}$). This means that the energy-reconstruction procedure described in Section~\ref{SubSect:MeshEnergy} is essential and that the artificial energies cannot be neglected. 

\begin{figure}
	\centering
	\includegraphics[scale=1]{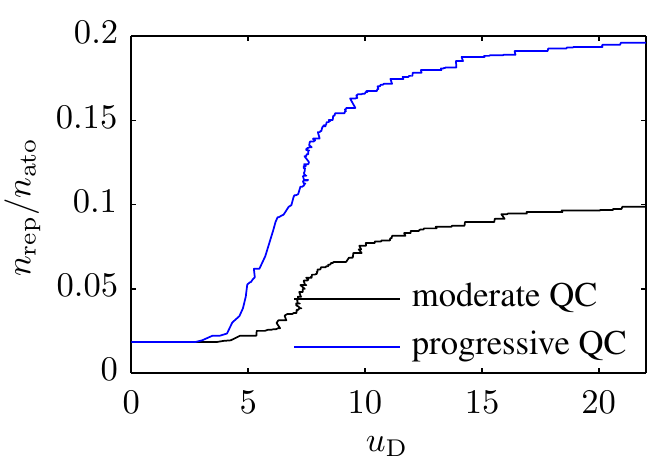}	
	\caption{Relative number of repatoms~$n_\mathrm{rep}/n_\mathrm{ato}$ in the L-shaped plate simulations as a function of~$u_\mathrm{D}$.}
	\label{SubSect:SimpleEx:Fig:4}
\end{figure}

\begin{figure}
	\centering
	\subfloat[$u_\mathrm{D} = 0$]{\includegraphics[scale=0.45]{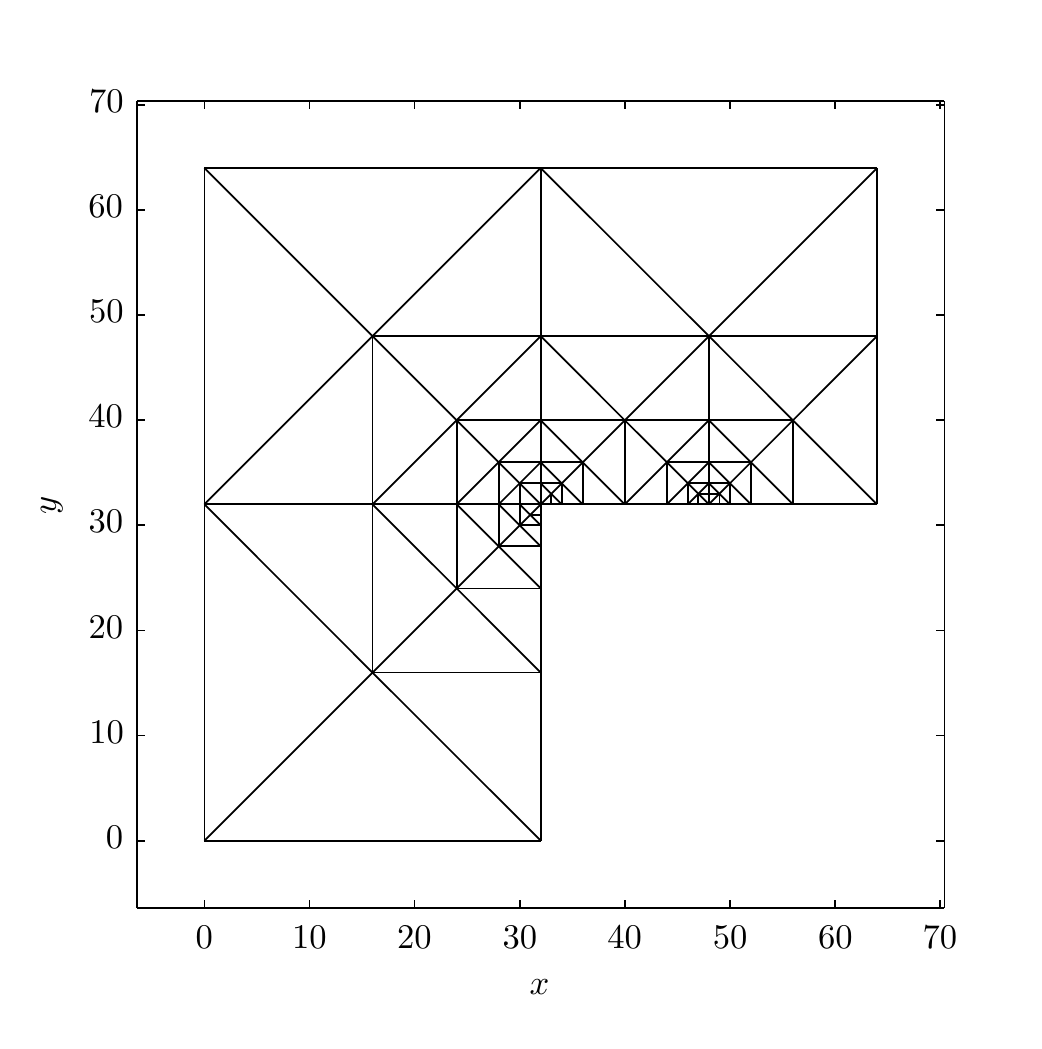}\label{SubSect:SimpleEx:Fig:3a}}\hspace{0.2em}
	\subfloat[$u_\mathrm{D} = 7$]{\includegraphics[scale=0.45]{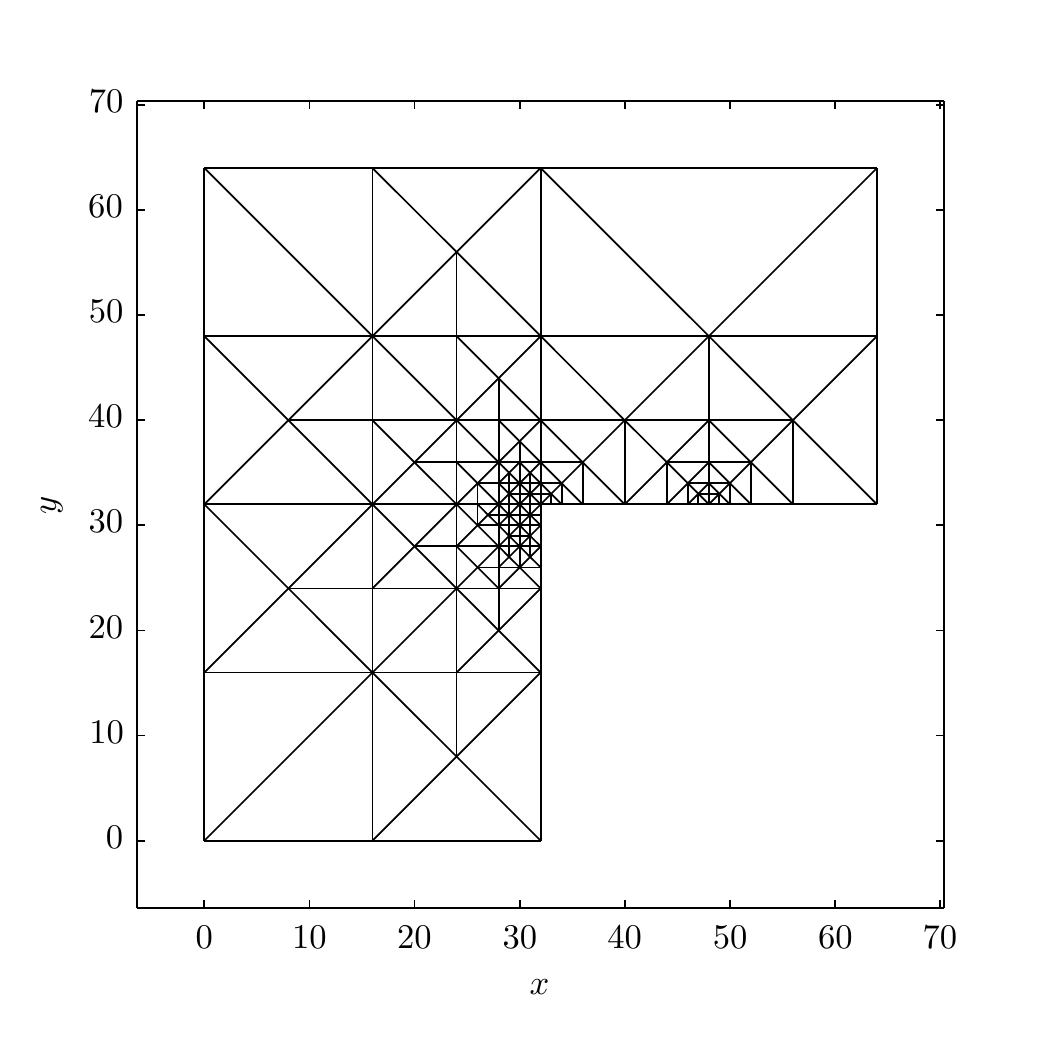}\label{SubSect:SimpleEx:Fig:3b}}\hspace{0.2em}
	\subfloat[$u_\mathrm{D} = 14$]{\includegraphics[scale=0.45]{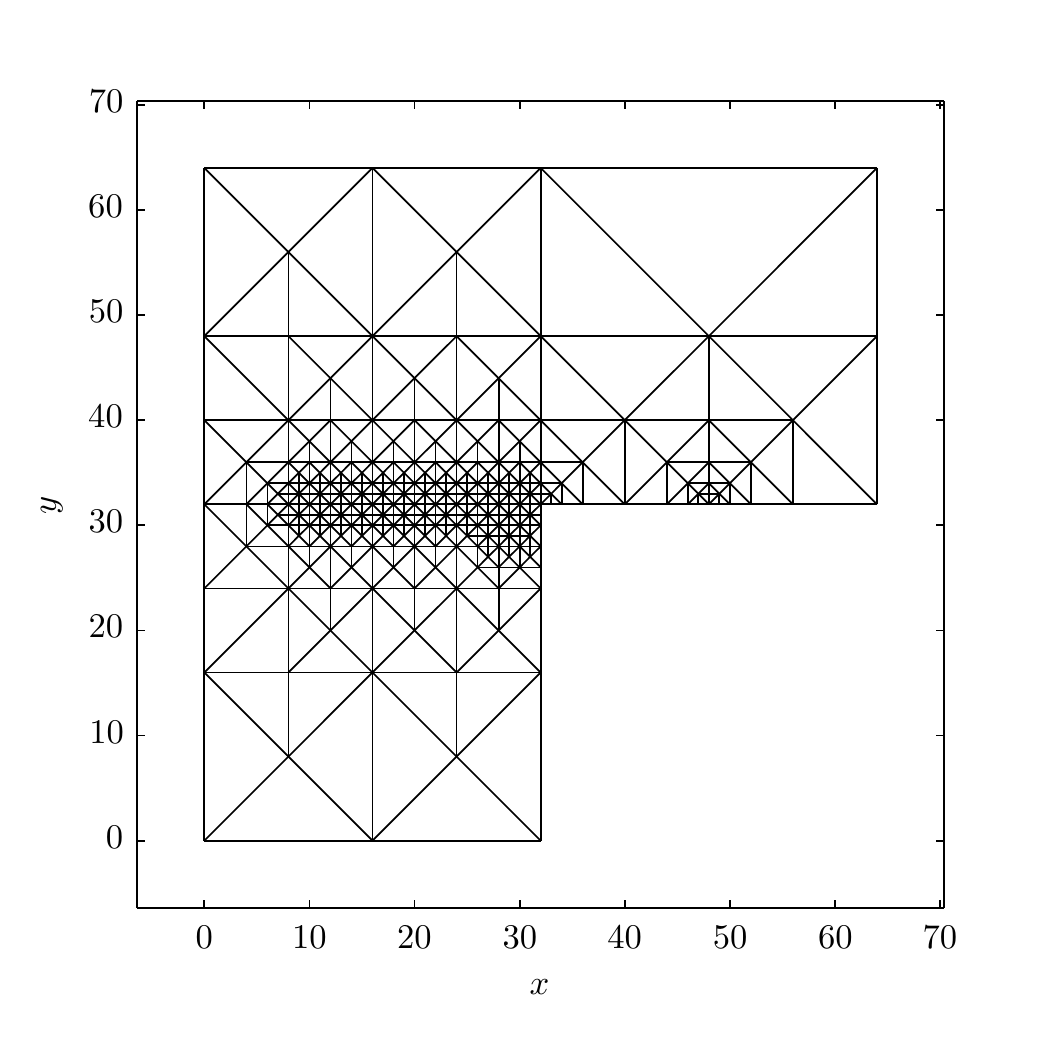}\label{SubSect:SimpleEx:Fig:3c}}\hspace{0.2em}
	\subfloat[$u_\mathrm{D} = 21$]{\includegraphics[scale=0.45]{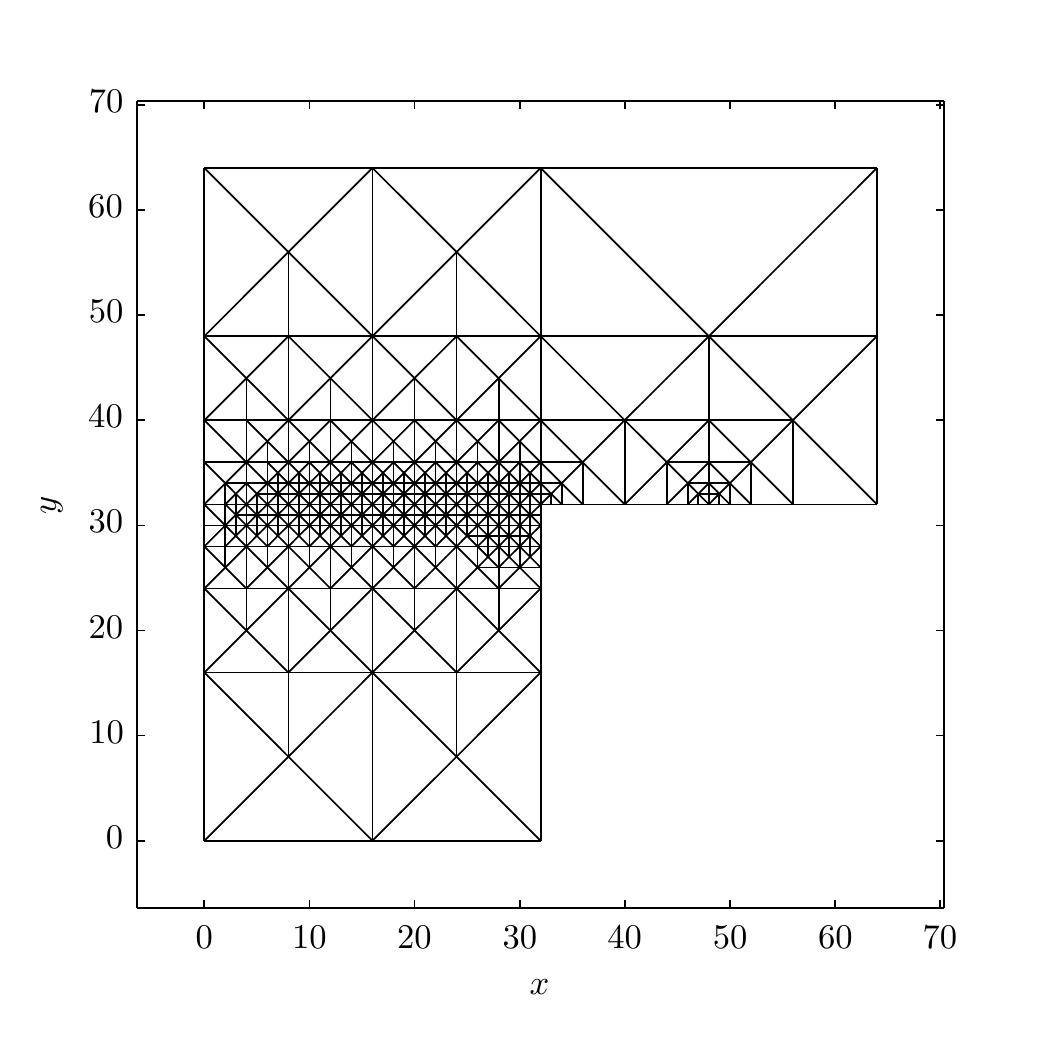}\label{SubSect:SimpleEx:Fig:3d}}\\	
	\subfloat[$u_\mathrm{D} = 0$]{\includegraphics[scale=0.45]{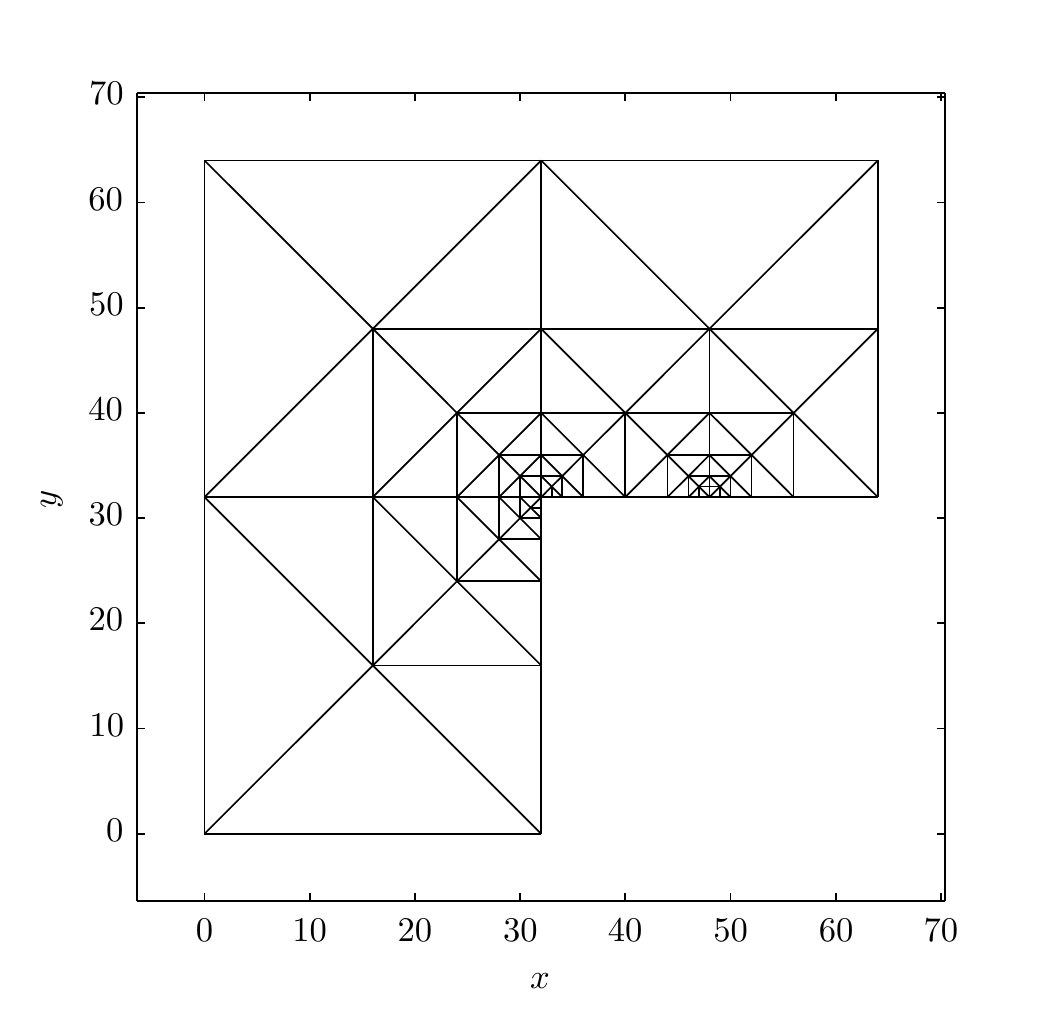}\label{SubSect:SimpleEx:Fig:3e}}\hspace{0.2em}
	\subfloat[$u_\mathrm{D} = 7$]{\includegraphics[scale=0.45]{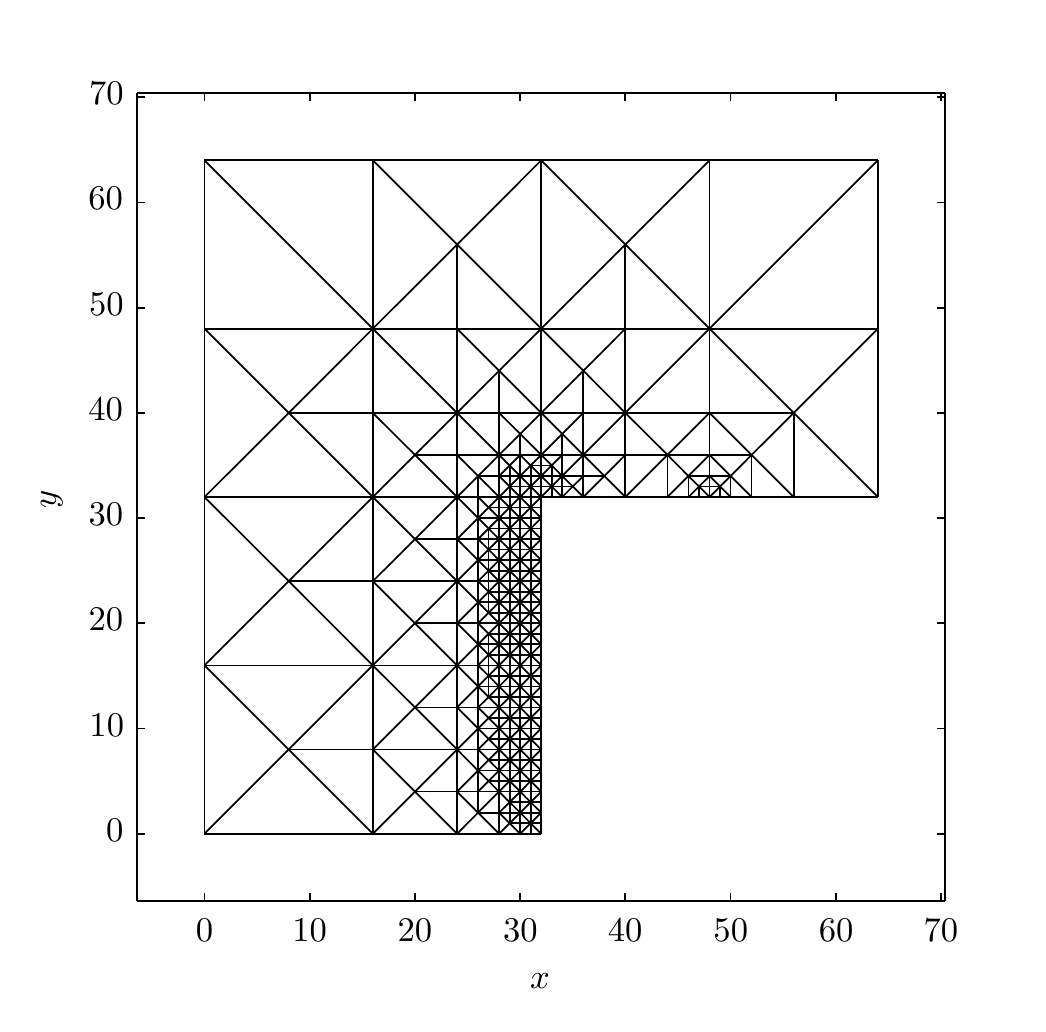}\label{SubSect:SimpleEx:Fig:3f}}\hspace{0.2em}
	\subfloat[$u_\mathrm{D} = 14$]{\includegraphics[scale=0.45]{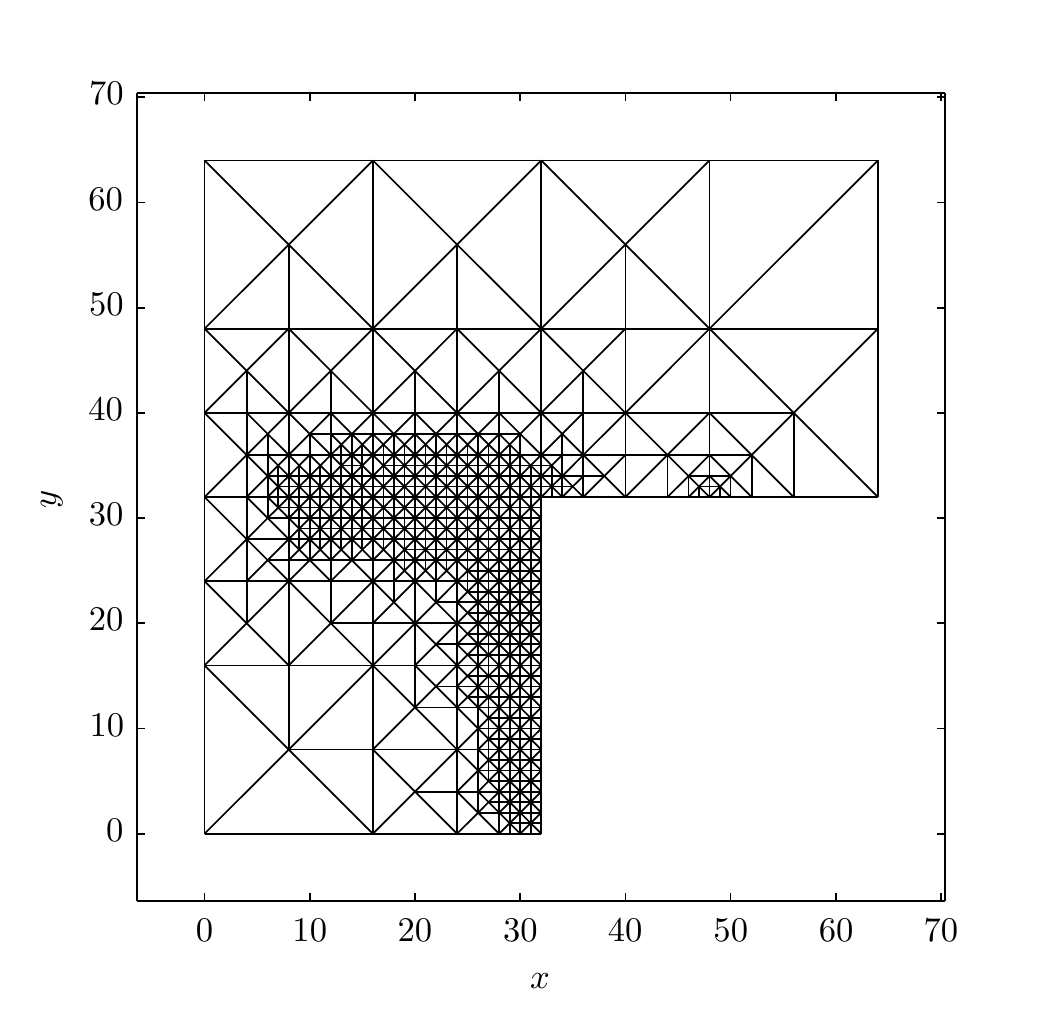}\label{SubSect:SimpleEx:Fig:3g}}\hspace{0.2em}
	\subfloat[$u_\mathrm{D} = 21$]{\includegraphics[scale=0.45]{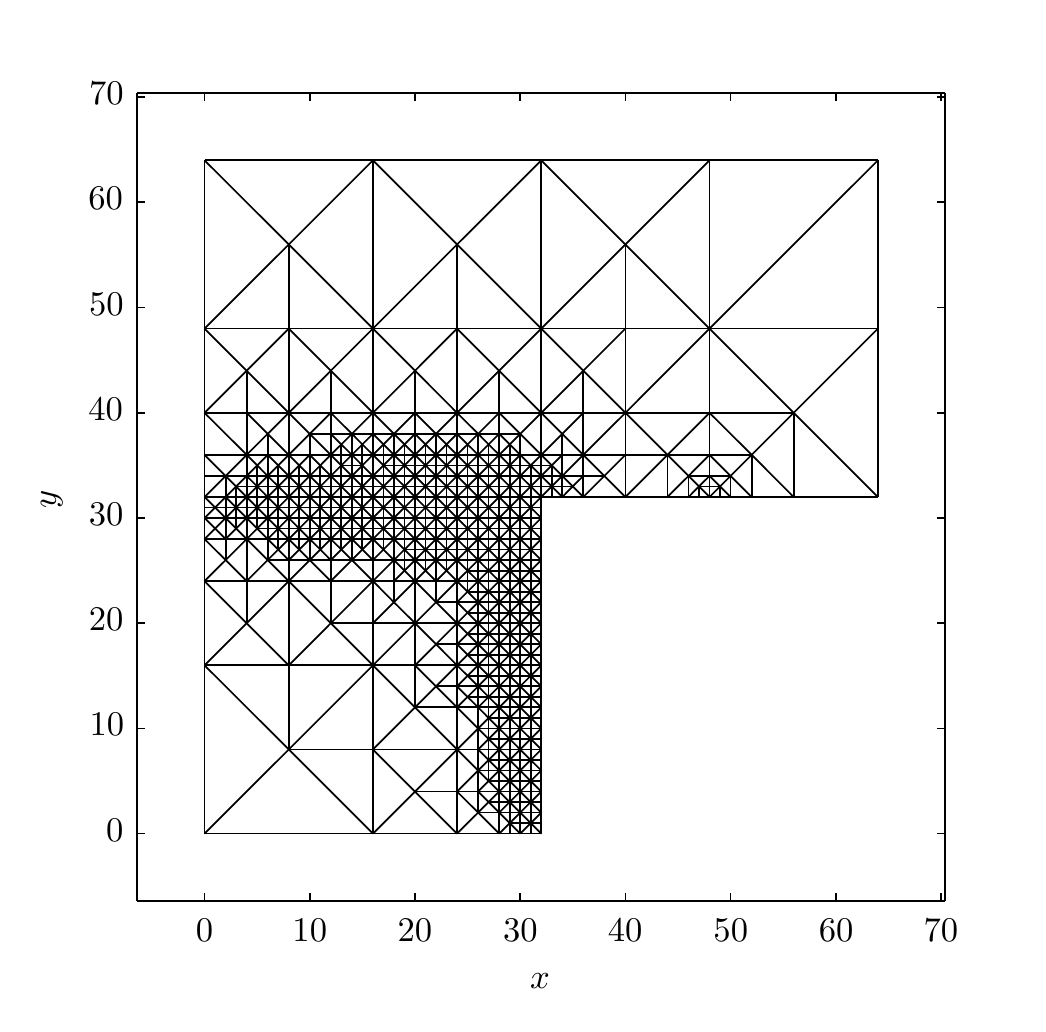}\label{SubSect:SimpleEx:Fig:3h}}
	\caption{Eight triangulations for the L-shaped plate test: (a)~-- (d) the moderate QC approach, $\theta = 0.5$, (e)~-- (h) the progressive QC approach, $\theta = 0.25$. For the relative number of repatoms corresponding to these meshes please refer to Fig.~\ref{SubSect:SimpleEx:Fig:4}.}
	\label{SubSect:SimpleEx:Fig:3}
\end{figure}

In Fig.~\ref{SubSect:SimpleEx:Fig:4}, the number of repatoms normalized by the total number of atoms (i.e.~$n_\mathrm{rep}(t_k)/n_\mathrm{ato}$) as a function of~$u_\mathrm{D}$ is shown. Because both curves are situated below~$0.2$, and the moderate QC below~$0.1$, appreciable computational savings are achieved. In particular, computing times were reduced by a factor of~$6.3$ for the moderate QC and~$3.5$ for the progressive QC compared to the full system.\footnote{Computing times are based on a Matlab implementation where computationally intensive parts (e.g. assembly of the gradients and Hessians) were coded in C++ and linked to the main code through mex files. Hence, due to this heterogeneity in the implementation, all computing times (and even their ratios) should be interpreted with great care, as they may not be representative. The simulations were performed using a personal computer with two cores (Intel Core 2 Duo E8400 @ 3.00GHz).} In terms of sampling interactions, the relative numbers of sampling interactions~$n_\mathrm{sam}^\mathrm{int}(t_k)/n_\mathrm{int}$ are slightly higher. Namely, below~$0.35$ and~$0.25$ for the progressive and moderate approach.

For completeness, Fig.~\ref{SubSect:SimpleEx:Fig:3} shows eight snapshots of the mesh evolution. Although both initial meshes~$\mathcal{T}_0$ are similar, different safety margins~$\theta$ cause the fully-resolved region of the progressive QC approach to be larger. Consequently, the obtained results are more accurate, but at the price that also regions far from the crack path are refined (e.g. along the~$\Gamma_2$ part of the boundary, cf. Fig.~\ref{SubSect:SimpleEx:Fig:3h}). The mesh of the moderate approach remains more localized, at the expense of a minor loss of accuracy.
%
%
\subsection{Antisymmetric Four-Point Bending Test}
\label{SubSect:ComplexEx}
\begin{figure}
	\centering
	\includegraphics[scale=1]{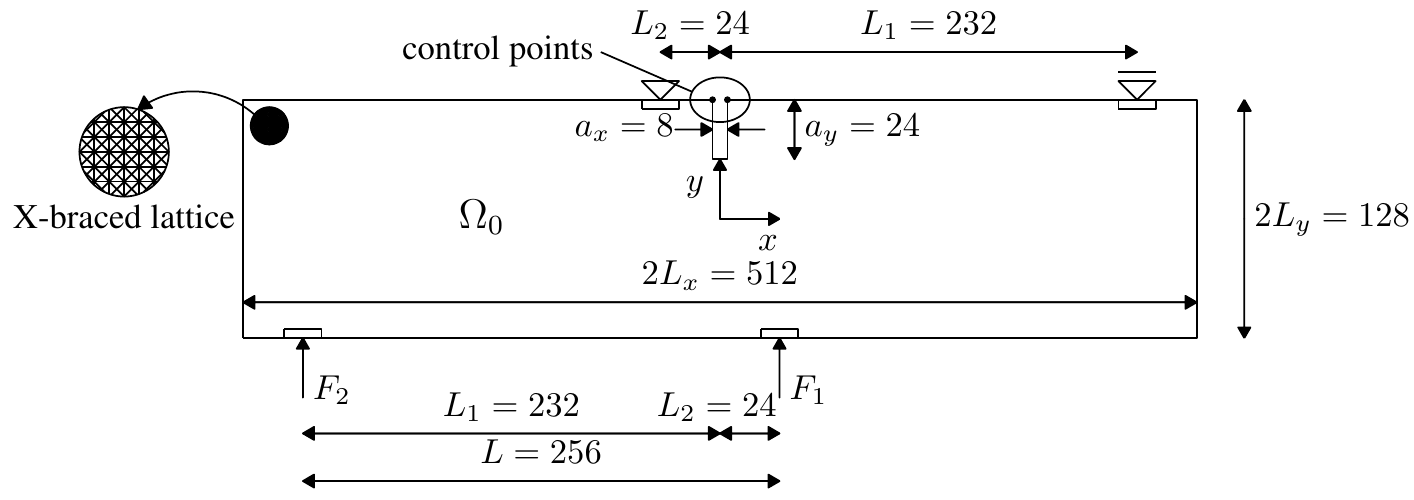}
	\caption{Sketch of the four-point bending test: geometry and boundary conditions.}
	\label{SubSect:ComplexEx:Fig:1}
\end{figure}
\begin{figure}
	\centering
	\subfloat[load control parameters]{\includegraphics[scale=1]{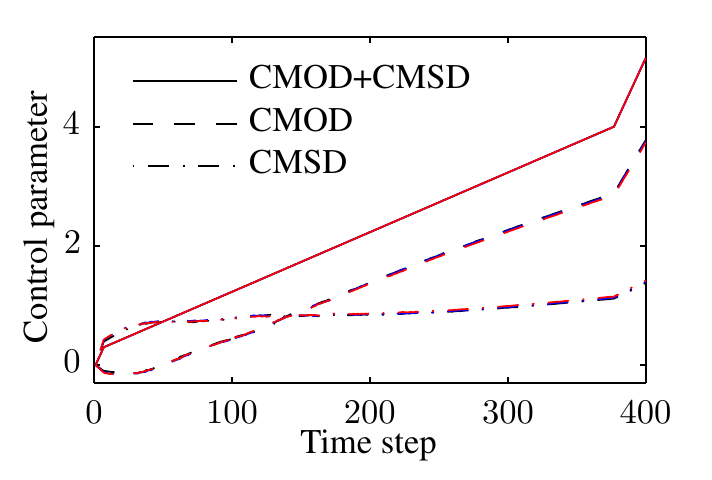}\label{SubSect:ComplexEx:Fig:2a}}\hspace{0.2em}
	\subfloat[$\bs{r}(t)$ for~$\mathrm{CMOD}+\mathrm{CMSD} = 5.5$]{\includegraphics[scale=1]{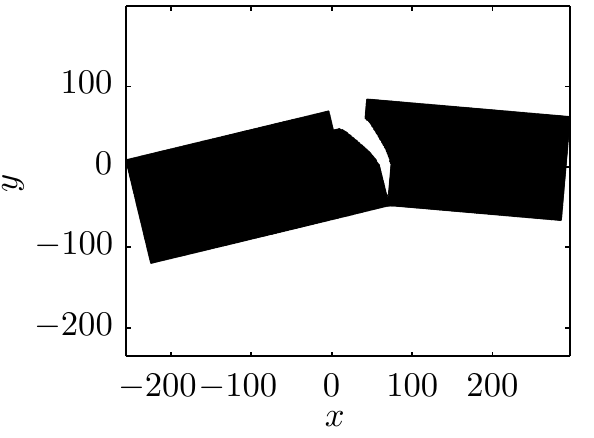}\label{SubSect:ComplexEx:Fig:2b}}
	\caption{Four-point bending test: (a)~evolution of CMOD, CMSD, and their sum (the applied loading program); colours: black~-- the moderate QC, $\theta = 0.5$, blue~-- the progressive QC, $\theta = 0.25$, red~-- full-lattice solution (note that they are almost indistinguishable). (b)~The deformed configuration for the full-lattice solution, displacements are magnified by a factor of~$10$.}
	\label{SubSect:ComplexEx:Fig:2}
\end{figure}

In the second example, a rectangular domain~$\Omega_0$ is exposed to antisymmetric four-point bending, cf. Fig.~\ref{SubSect:ComplexEx:Fig:1} and, e.g., \cite{SchlangenThesis}. The homogeneous body is pre-notched from the top edge to initiate a crack, and stiffened locally where prescribed displacements and forces are applied (again, the Young's modulus is~$1000$ times larger than elsewhere and the limit elastic strain~$\varepsilon_0$ is infinite to prevent any damage evolution). The lattice spacing is of a unit length in both directions. The entire specimen consists of~$66,009$ atoms connected by~$262,040$ interactions. The (vertical) forces~$F_1$ and~$F_2$ are prescribed as, cf. Fig.~\ref{SubSect:ComplexEx:Fig:1},
\begin{equation}
F_1 = \frac{L_1}{L}\lambda, \quad F_2 = \frac{L_2}{L}\lambda,
\label{SubSect:ComplexEx:Eq1}
\end{equation}
where~$\lambda$ is the additional parameter used for indirect displacement solution control, cf. Eq.~\eqref{SubSect:SimpleEx:Eq2} and the discussion on it. In contrast to the previous example, the sum of CMOD and CMSD is used to control the simulation (recall Eq.~\eqref{SubSect:SimpleEx:Eq2} where~$\bs{c}_\mathrm{o} \neq \bs{0}$ and~$\bs{c}_\mathrm{s} \neq \bs{0}$). This combination of the two measures is required because of the following reasons. Initially, the CMOD is close to zero or even negative whereas the CMSD drives the evolution. In the later stages, however, the CMSD is constant while CMOD parametrizes the process. Their sum, therefore, naturally switches between the two approaches, see also Fig.~\ref{SubSect:ComplexEx:Fig:2a}. Due to a higher brittleness compared to the previous example, cf. Tab.~\ref{Sect:Examples:Tab:1}, two loading rates for $\mathrm{CMOD}+\mathrm{CMSD}$, as specified in Fig.~\ref{SubSect:ComplexEx:Fig:2a}, are used. 

The numerical example is studied again for fully resolved system and the two QC approaches: the moderate QC, $\theta = 0.5$, and the progressive QC, $\theta = 0.25$. In order to achieve a higher accuracy using the progressive approach, a globally fine initial mesh is used, in which the maximum triangle edge length is restricted to~$16$ lattice spacings. For the moderate approach, the mesh is as coarse as possible to describe the specimen geometry by a right-angled triangulation. The initial meshes are the top triangulations in Fig.~\ref{SubSect:ComplexEx:Fig:5} (Figs.~\ref{SubSect:ComplexEx:Fig:5a} and~\ref{SubSect:ComplexEx:Fig:5e}).
\begin{figure}
	\centering
	\subfloat[crack paths]{\includegraphics[scale=1]{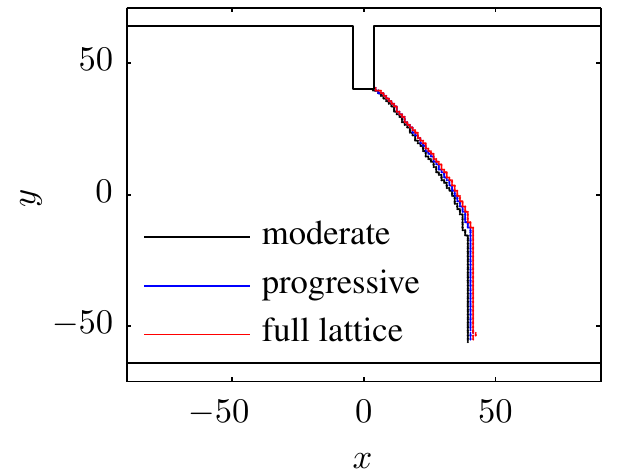}\label{SubSect:ComplexEx:Fig:3a}}\hspace{1em}
	\subfloat[force-opening diagram]{\includegraphics[scale=1]{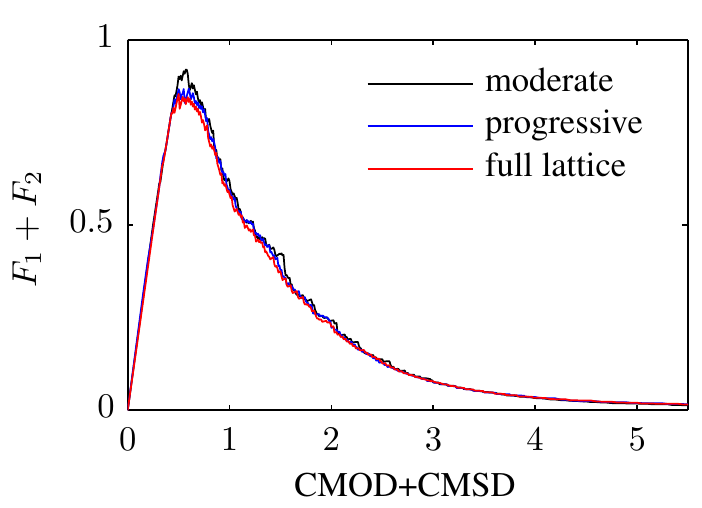}\label{SubSect:ComplexEx:Fig:3b}}
	\caption{Four-point bending test: (a)~crack paths, and~(b) force-opening diagram, i.e.~$\lambda = F_1 + F_2$ versus~$\mathrm{CMOD}+\mathrm{CMSD}$.}
	\label{SubSect:ComplexEx:Fig:3}
\end{figure}

The deformed configuration predicted by the full-lattice solution at~$\mathrm{CMOD} + \mathrm{CMSD} = 5.5$ is presented in Fig.~\ref{SubSect:ComplexEx:Fig:2b}. In qualitative accordance with experimental data, see e.g.~\cite{SchlangenThesis}, Section~4.1, the crack path initiates at the right bottom corner of the notch, subsequently curves downwards and then approaches the bottom part of the boundary to the right side of the force~$F_1$. The crack paths predicted by the full-lattice simulation and both adaptive QC schemes are presented on the undeformed configuration in Fig.~\ref{SubSect:ComplexEx:Fig:3a}. Here we notice that the results are almost identical. The total applied force~$\lambda = F_1 + F_2$, recall Eq.~\eqref{SubSect:ComplexEx:Eq1}, is plotted in Fig.~\ref{SubSect:ComplexEx:Fig:3b} against~$\mathrm{CMOD}+\mathrm{CMSD}$. Although the initial triangulations differ significantly (cf. Figs.~\ref{SubSect:ComplexEx:Fig:5a} and~\ref{SubSect:ComplexEx:Fig:5e}), the initial slopes are practically identical. As the moderate QC refines later and less extensively, the peak force is overestimated by the moderate QC compared to the full-lattice solution by approximately~$10\,\%$, whereas the progressive QC is overall accurate. In the post-peak region, all curves are practically identical again.
\begin{figure}
	\centering
	\subfloat[energy evolutions]{\includegraphics[scale=1]{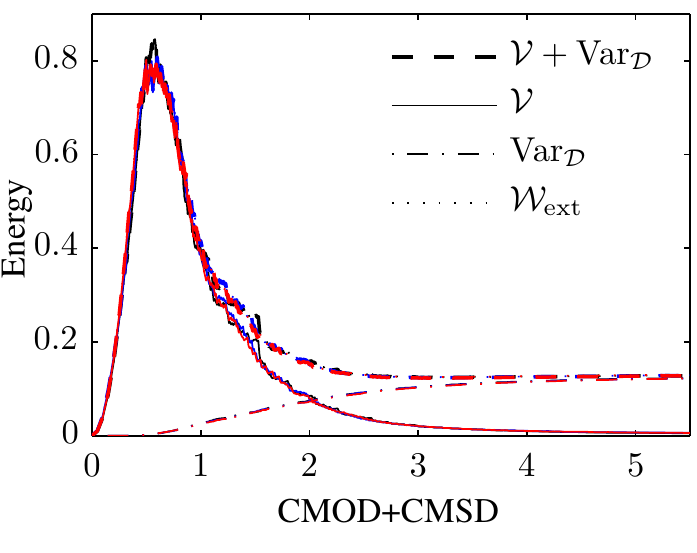}\label{SubSect:ComplexEx:Fig:4a}}\hfill
	\subfloat[energy components for moderate approach]{\includegraphics[scale=1]{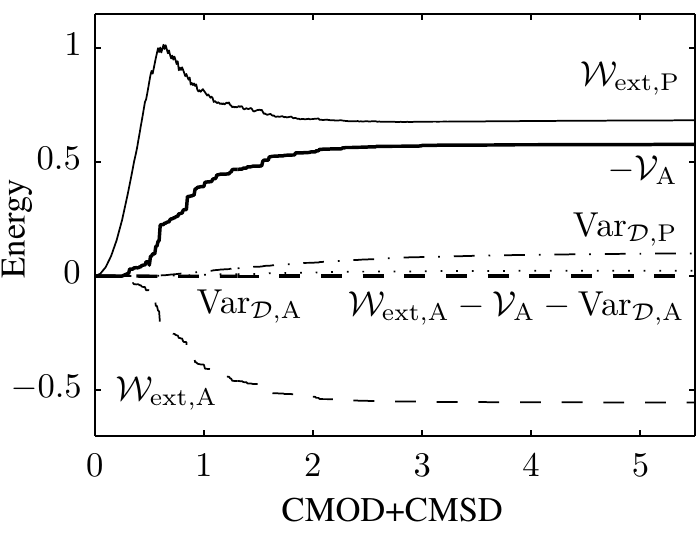}\label{SubSect:ComplexEx:Fig:4b}}
	\caption{Four-point bending test. (a)~Energy evolutions (black~-- the moderate QC approach, $\theta = 0.5$, blue~-- the progressive QC approach, $\theta = 0.25$, red~-- full-lattice solution). (b)~Energies exchanged during mesh refinement for the moderate QC approach, see Section~\ref{SubSect:MeshEnergy}.}
	\label{SubSect:ComplexEx:Fig:4}
\end{figure}

\begin{figure}
	\centering
	\includegraphics[scale=1]{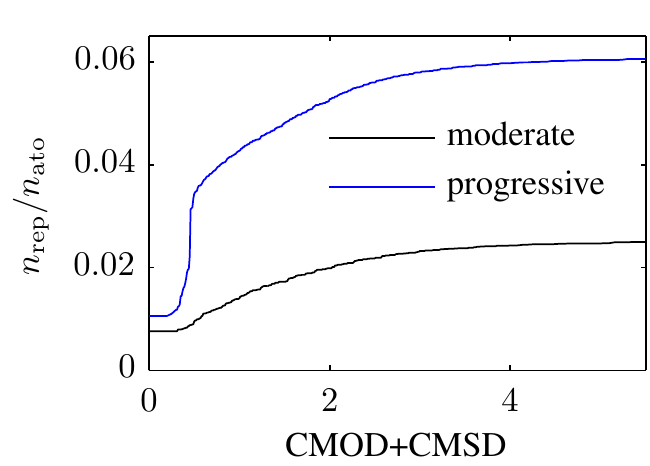}
	\caption{Four-point bending test: the relative number of repatoms~$n_\mathrm{rep}/n_\mathrm{ato}$ as a function of~$\mathrm{CMOD}+\mathrm{CMSD}$. Eight chosen triangulations corresponding to~$\mathrm{CMOD}+\mathrm{CMSD} = 0$, $1$, $2$, and~$5.5$ are presented in Fig.~\ref{SubSect:ComplexEx:Fig:5}.}
	\label{SubSect:ComplexEx:Fig:6}
\end{figure}

\begin{figure}
	\centering
	\subfloat[moderate, $\mathrm{CMOD}+\mathrm{CMSD} = 0$]{\includegraphics[scale=0.425]{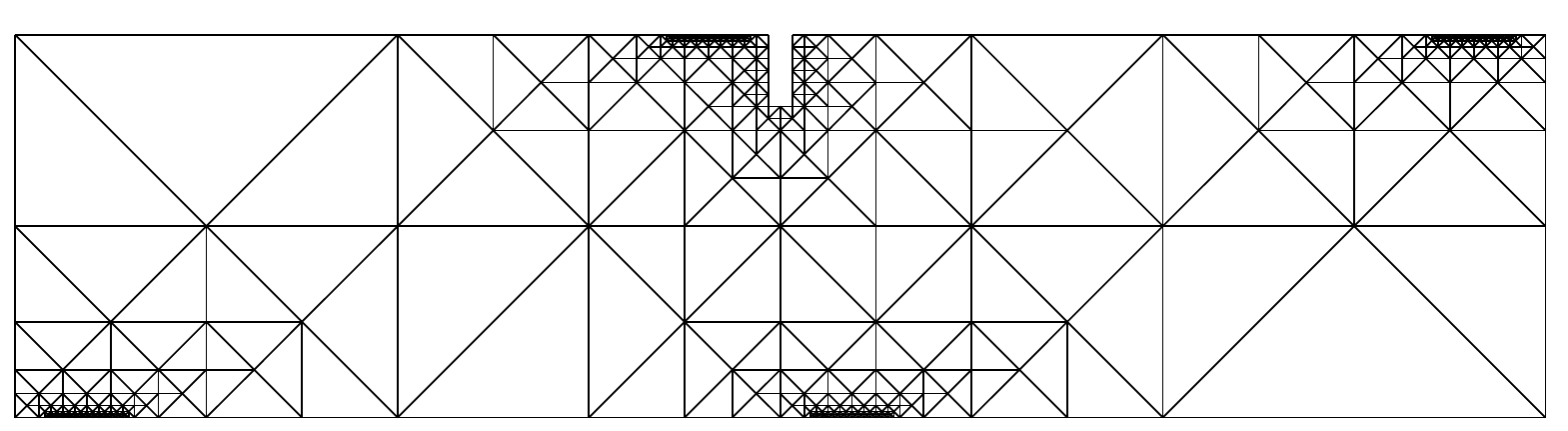}\label{SubSect:ComplexEx:Fig:5a}}\hspace{0.2em}
	\subfloat[progressive, $\mathrm{CMOD}+\mathrm{CMSD} = 0$]{\includegraphics[scale=0.425]{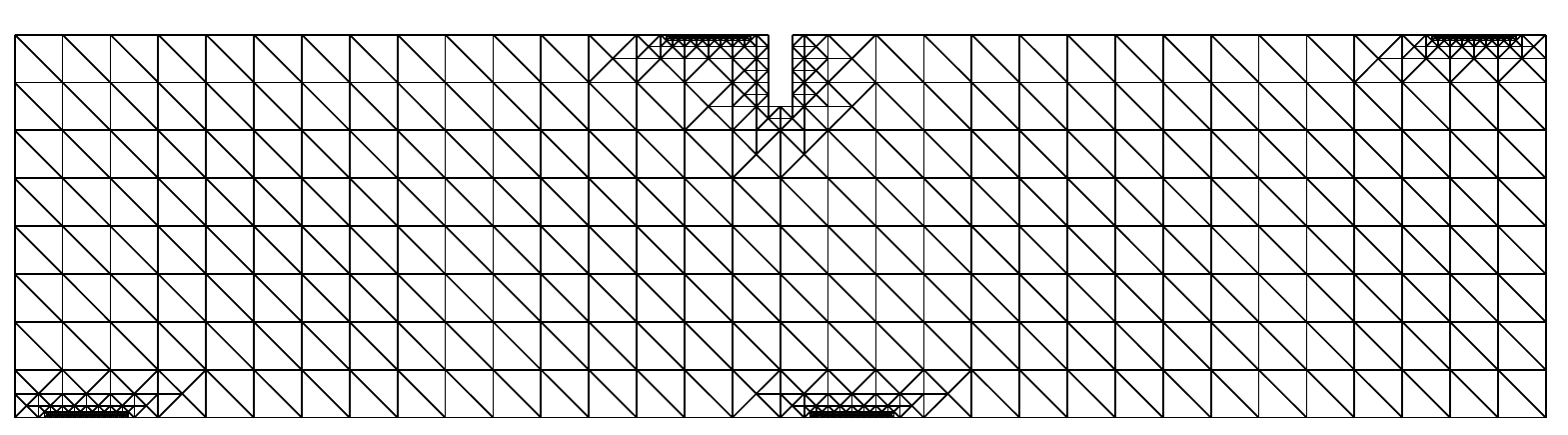}\label{SubSect:ComplexEx:Fig:5e}}\\
	\subfloat[moderate, $\mathrm{CMOD}+\mathrm{CMSD} = 1$]{\includegraphics[scale=0.425]{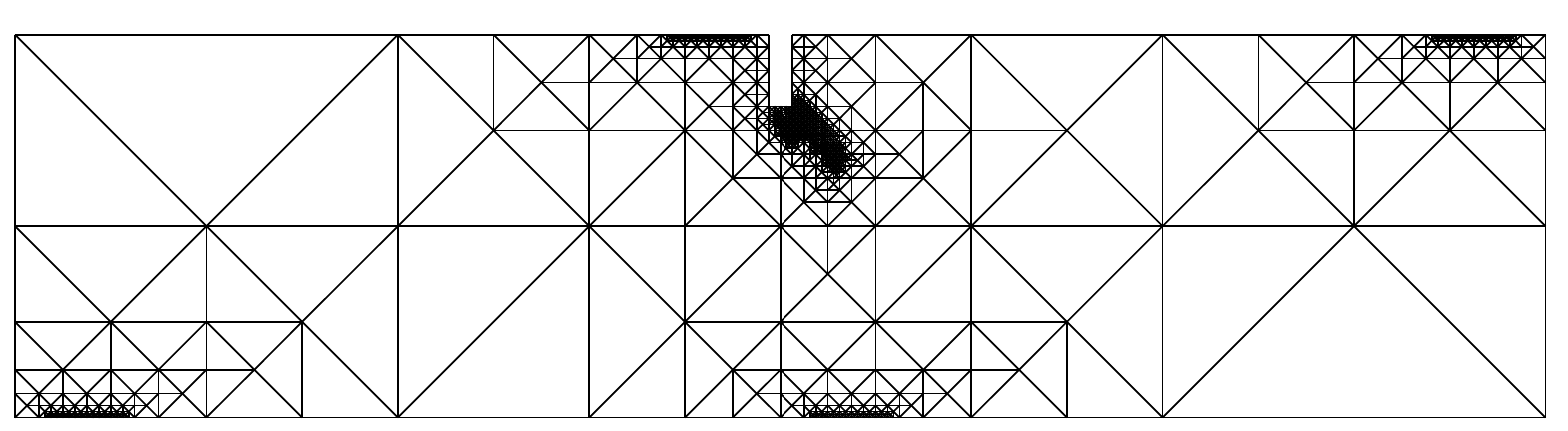}\label{SubSect:ComplexEx:Fig:5b}}\hspace{0.2em}
	\subfloat[progressive, $\mathrm{CMOD}+\mathrm{CMSD} = 1$]{\includegraphics[scale=0.425]{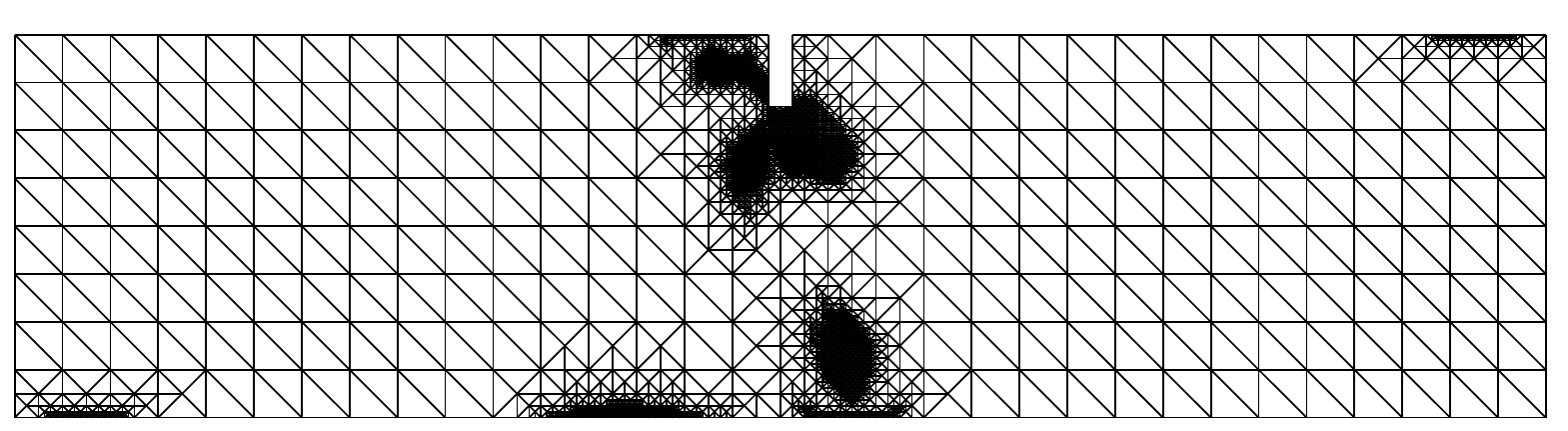}\label{SubSect:ComplexEx:Fig:5f}}\\
	\subfloat[moderate, $\mathrm{CMOD}+\mathrm{CMSD} = 2$]{\includegraphics[scale=0.425]{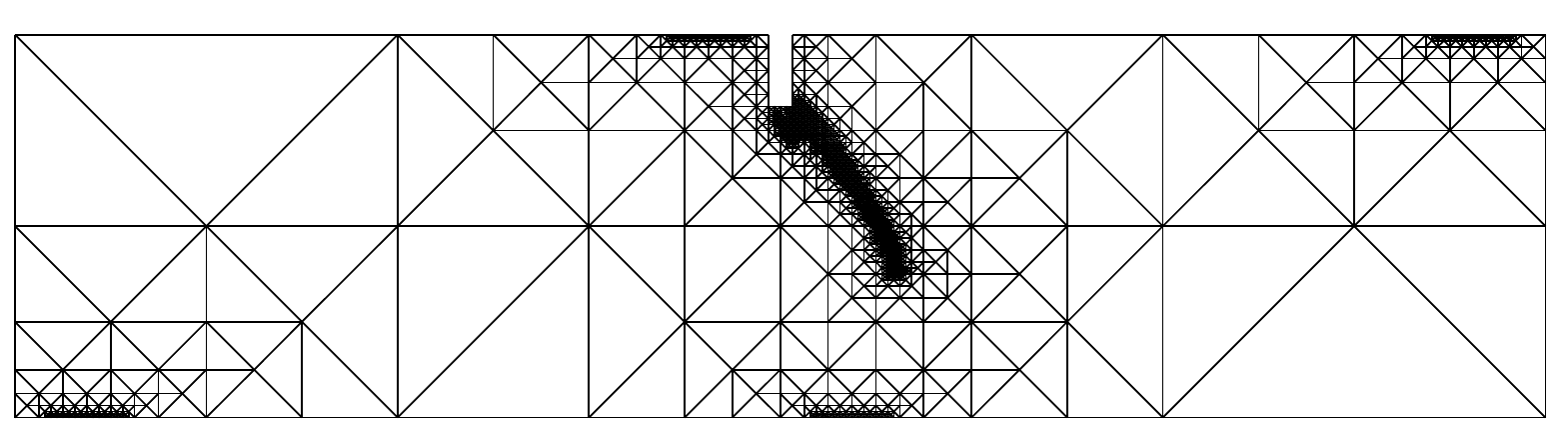}\label{SubSect:ComplexEx:Fig:5c}}\hspace{0.2em}
	\subfloat[progressive, $\mathrm{CMOD}+\mathrm{CMSD} = 2$]{\includegraphics[scale=0.425]{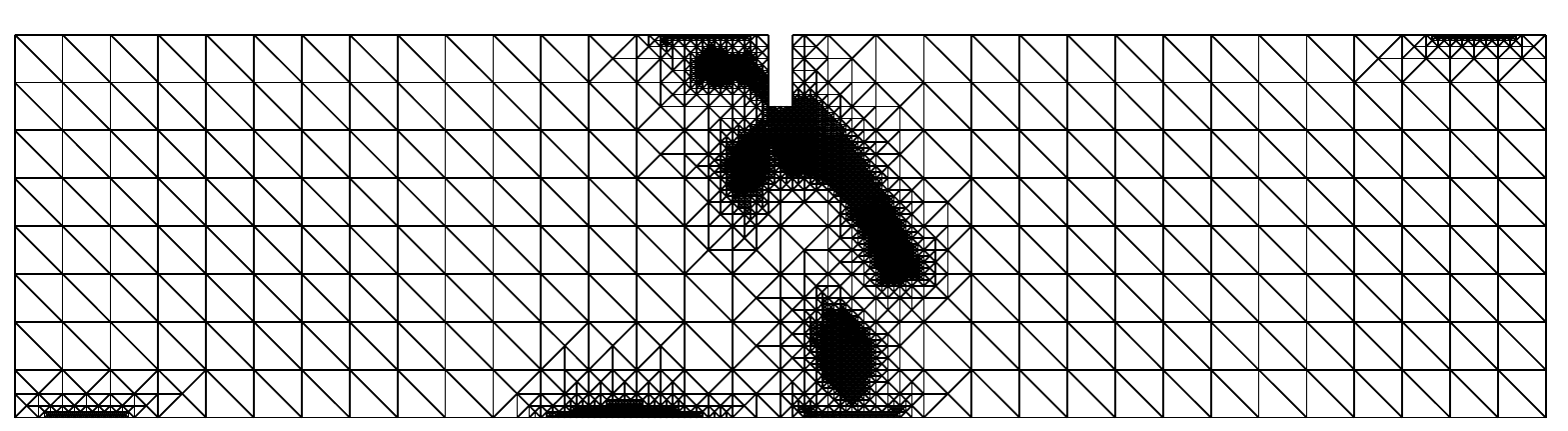}\label{SubSect:ComplexEx:Fig:5g}}\\
	\subfloat[moderate, $\mathrm{CMOD}+\mathrm{CMSD} = 5.5$]{\includegraphics[scale=0.425]{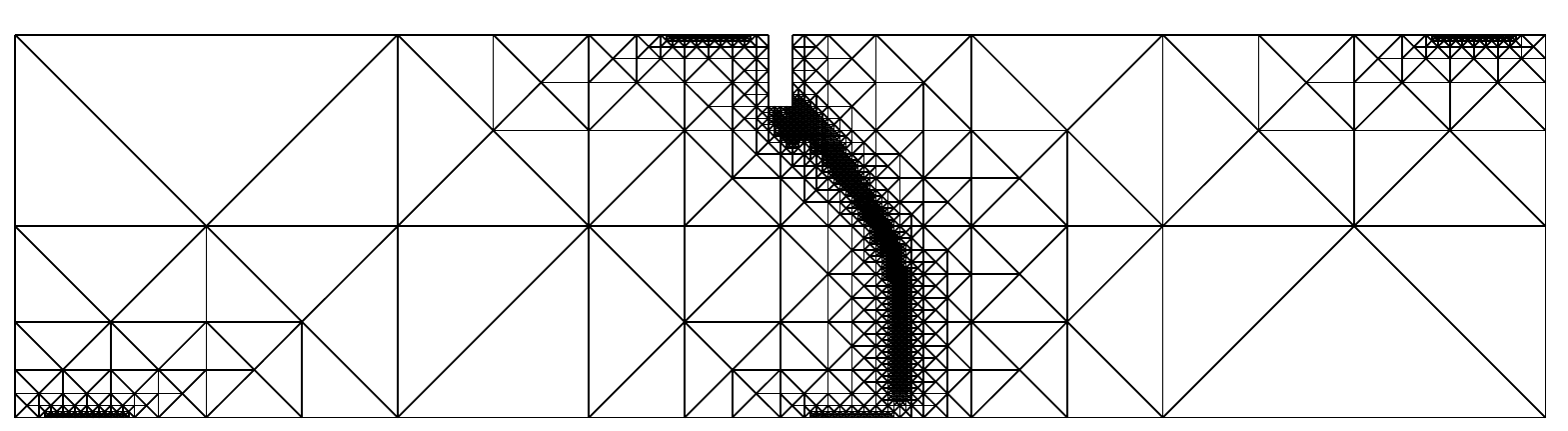}\label{SubSect:ComplexEx:Fig:5d}}\hspace{0.2em}
	\subfloat[progressive, $\mathrm{CMOD}+\mathrm{CMSD} = 5.5$]{\includegraphics[scale=0.425]{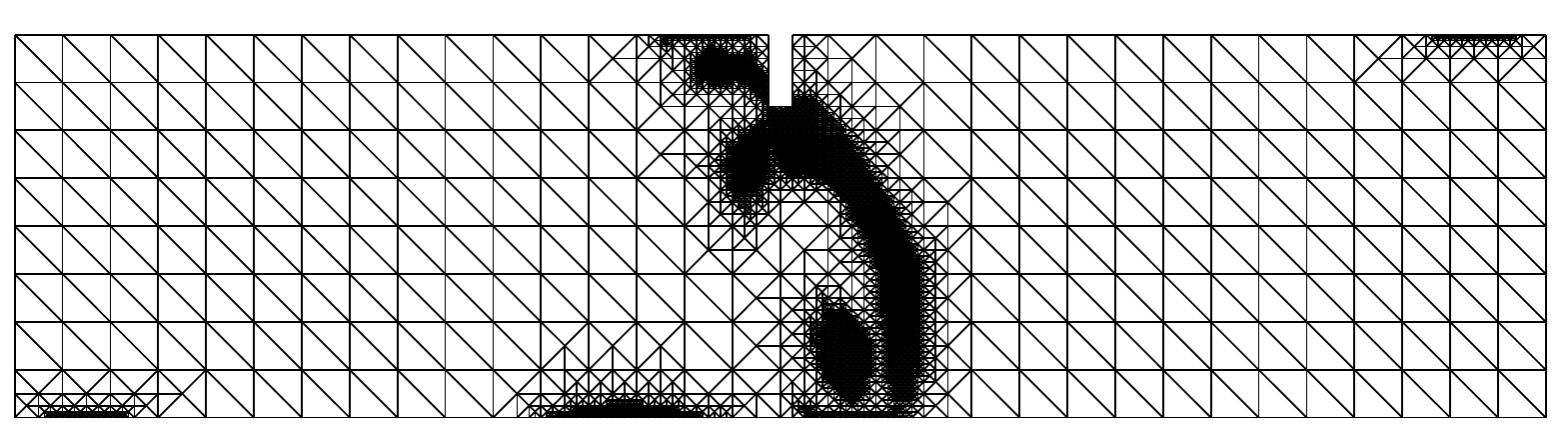}\label{SubSect:ComplexEx:Fig:5h}}
	\caption{Eight triangulations for four-point bending test: (a), (c), (e), (g) correspond to the moderate QC approach, $\theta = 0.5$, and~(b), (d), (f), (h) to the progressive QC approach, $\theta = 0.25$. The relative numbers of repatoms are shown in Fig.~\ref{SubSect:ComplexEx:Fig:6}.}
	\label{SubSect:ComplexEx:Fig:5}
\end{figure}

The energy evolution paths corresponding to all approaches are presented in Fig.~\ref{SubSect:ComplexEx:Fig:4a}. From there it may be concluded that the results match well. Moreover, we see that all solutions satisfy the energy balance~\eqref{E} along the entire loading path. In Fig.~\ref{SubSect:ComplexEx:Fig:4b}, substantial energy exchanges (i.e. artificial energies) due to mesh refinement can again be observed (similar to the first numerical example). Because the size of the fully-resolved domain is small compared to the entire domain, both adaptive QC approaches achieve a substantial computational gain; the corresponding computing times were reduced by factors of~$27.9$ (for the moderate QC) and~$11.1$ (for the progressive QC) compared to the full system. This is also supported by Fig.~\ref{SubSect:ComplexEx:Fig:6} in which the relative numbers of repatoms are presented. The ratio remains below~$0.065$ and even below~$0.025$ for the moderate refinement strategy. In the case of the relative numbers of sampling interactions, the ratios remain below~$0.12$ and~$0.06$ for the progressive and moderate approaches. Note also that the number of repatoms increases rapidly near the peak load for the progressive approach whereas it develops more gradually for the moderate one.

Finally, in Fig.~\ref{SubSect:ComplexEx:Fig:5} several snapshots that capture the evolution of the triangulations~$\mathcal{T}_k$ are presented. It can again be noticed that the progressive approach refines until quite far from the crack tip, whereas the fully-refined region in the moderate QC remains localized.
%
%
\section{Summary and Conclusions}
\label{Sect:Conclusion}
In this contribution, we have developed an energy-based dissipative QC approach for regular lattice networks with damage and fracture. The study shows that the efficiency of the QC methodology applies also to brittle phenomena, and that together with an adaptive refinement strategy it provides a powerful tool to predict crack propagation in lattice networks. The main results can be summarized as follows:
\begin{enumerate}
	\item The general variational formulation for rate-independent processes by Mielke and Roub\'{i}\v{c}ek~\cite{MieRou:2015} was rephrased for the case of lattice networks with damage.

	\item The two standard QC steps, interpolation and summation, were revisited from an adaptive point of view. For the interpolation, meshes with right-angled triangles were used because of their
		\begin{itemize}
			\item[(i)] ability to naturally refine to the fully-resolved underlying lattice

			\item[(ii)] binary-tree structure that allows for fast and efficient data transfer
			
			\item[(iii)] significant reduction of the summation part of the QC error.
		\end{itemize}
	
	\item To determine the location of the critical region, a heuristic marking strategy with a variable parameter that controls the accuracy of the simulation was proposed.
	
	\item The mesh refinement procedure was discussed from an energetic standpoint, and the significance of the reconstruction procedure with respect to the energy consistency was shown.

	\item The numerical examples demonstrated that the introduced marking strategy is capable of satisfactorily predicting the evolution of the crack path and the load-displacement response, especially in the post-peak region. Solutions obtained using indirect load displacement control satisfied the energy equality condition.
\end{enumerate}

Let us note that as the crack tip propagates throughout the body, it would be convenient to include besides the mesh refinement ahead of it also mesh coarsening in its wake. Furthermore, instead of the proposed heuristic marking strategy, techniques such as goal-oriented error estimators may be implemented to improve further the performance of the adaptive QC. Both aspects enjoy our current interest and will be reported separately.
%
\appendix
%
%
\section{Derivation of the Dissipation Distance}
\label{Sect:A}
This Appendix provides the details on the dissipation function~$D(\omega)$ needed in the dissipation distance~$\mathcal{D}$ of Eq.~\eqref{SubSect:DissLatt:Eq:4} for the constitutive law with exponential softening shown in Fig.~\ref{Sect:Examples:Fig:1}, cf. also Eq.~\eqref{Examples:Eq:2}. Because damage evolves only under tension, recall Section~\ref{SubSect:Energies}, we assume in the remainder of this section that~$\widehat{r} > r_0$ for the ease of notation. Recall that~$r_0$, in accordance with Eq.~\eqref{SubSect:DissLatt:Eq:1a}, denotes the initial length of the interaction, and~$\widehat{r}$~its (admissible) deformed length. In order to make the interaction behaviour independent of the sectional area and the initial length of the bond, we introduce interaction strain, $\widehat{\varepsilon} = (\widehat{r} - r_0)/r_0$, and the interaction stress (the superscripts~$\alpha\beta$ are dropped for the sake of brevity)
\begin{equation}
\sigma = \frac{N}{A} = (1-\widehat{\omega})\frac{\phi'(\widehat{r})}{A} = (1-\widehat{\omega})E\widehat{\varepsilon},
\label{Sect:Examples:Eq:1}
\end{equation}
where the normal force of the interaction is denoted as~$N = \frac{\mathrm{d}}{\mathrm{d}\,\widehat{r}}\,\widetilde{\pi}^k(\widehat{r},\widehat{\omega};\bs{q}(t_{k-1}))=(1-\widehat{\omega})\phi'(\widehat{r})$, and where we have used~$\phi'(\widehat{r})=\frac{\mathrm{d}}{\mathrm{d}\,\widehat{r}}\,\phi(\widehat{r})$. Remind that~$E$~is the Young's modulus, $A$~the cross-sectional area, and~$\widehat{\omega}$ associated (admissible) damage variable. The pair potential~$\phi$ from Eq.~\eqref{Examples:Eq:1} can be rewritten as
\begin{equation}
\phi(\widehat{r}) = \frac{1}{2}EAr_0\left(\widehat{\varepsilon}(\widehat{r})\right)^2,
\label{Sect:Examples:Eq:2}
\end{equation}
where we have emphasized that the admissible strain~$\widehat{\varepsilon}$ is a function of the admissible length~$\widehat{r}$. To construct a constitutive model that under monotonic damage evolution displays a specific stress-strain response, say
\begin{equation}
\sigma = s(\widehat{\varepsilon}),
\label{pstress}
\end{equation}
where~$s(\bullet)$ is a given target softening function (recall Eq.~\eqref{Examples:Eq:2} and Fig.~\ref{Sect:Examples:Fig:1}), the damage variable is considered as a function of the current strain. Using the constitutive relation in Eq.~\eqref{Sect:Examples:Eq:1} while employing~\eqref{pstress}, one obtains
\begin{equation}
\widehat{\omega} = 1-\frac{s(\widehat{\varepsilon})}{E\widehat{\varepsilon}}=:g(\widehat{\varepsilon}),
\label{damagelaw}
\end{equation}
i.e. the damage variable as a function of the strain~$\widehat{\varepsilon}$. In the interval of growing damage, the function~$g(\widehat{\varepsilon})$ is invertible, providing
\begin{equation}
\widehat{\varepsilon} = g^{-1}(\widehat{\omega}).
\label{damageInversion}
\end{equation}
Eq.~\eqref{damageInversion} substituted into~\eqref{damaging} (where~$\phi$ is now considered as a function of~$\widehat{\varepsilon}$ rather than~$\widehat{r}$ according to~\eqref{Sect:Examples:Eq:2}) provides\footnote{Note that although we use hatted variables that indicate arbitrary admissible configurations, the constitutive law is actually ensured by the minimization with respect to~$\widehat{\omega}$; recall the first-order optimality conditions in Eqs.~\eqref{elastic}~-- \eqref{damaged}, where the minimizer was denoted as~$\mathring{\omega}^{\alpha\beta}$, cf. also Eqs.~\eqref{SubSect:FullSol:Eq:3} and~\eqref{SubSect:FullSol:Eq:4}. Using hats in Eq.~\eqref{SubSect:FullSol:Eq:8} is therefore a slight abuse of notation as~$\widehat{\omega}$ is not entirely arbitrary.}
\begin{equation}
D'(\widehat{\omega}) = \phi(\widehat{\varepsilon}) = \frac{1}{2}EAr_0\left(g^{-1}(\widehat{\omega})\right)^2.
\label{SubSect:FullSol:Eq:8}
\end{equation}
Integrating this relation yields
\begin{equation}
D(\widehat{\omega}) = \frac{EAr_0}{2}\int_{0}^{\widehat{\omega}}\left(g^{-1}(\eta)\right)^2\,\mathrm{d}\eta,
\label{SubSect:FullSol:Eq:9}
\end{equation}
that can be expressed in a closed form for some special cases such as linear softening, cf.~\cite{JiZe:Damage}, Eq.~(85). Before proceeding, let us note that from the computational point of view, the knowledge of~$D$ is required only for the verification of the energy balance~\eqref{E}, and that in accordance with Eqs.~\eqref{elastic}~-- \eqref{damaged} the solution actually requires only~$D'$. Hence, for the computational purposes, the definition in Eq.~\eqref{damagelaw} is fully sufficient,\footnote{It may be also clear at this stage that solving for~$\mathring{\omega}$ in~\eqref{SubSect:FullSol:Eq:3} reduces to function evaluation in~\eqref{damagelaw}, where the damage is computed as a function of strain (upon accounting for irreversibility).} whereas integration of~$D$ in~\eqref{SubSect:FullSol:Eq:9} can be carried out numerically.

For the exponential softening law defined in Eq.~\eqref{Examples:Eq:2}, the expression~\eqref{damagelaw} attains the following form
\begin{figure}
	\centering
	\subfloat[$\omega = g(\varepsilon)$]{\includegraphics[scale=1]{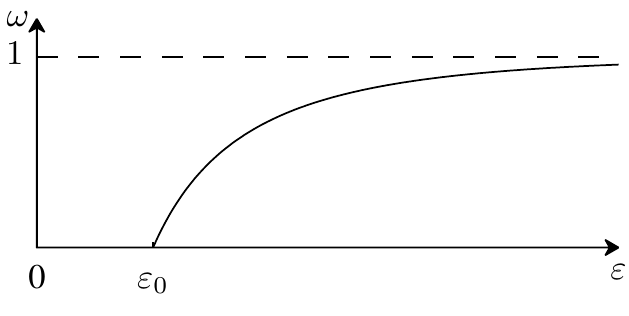}\label{Sect:A:Fig:1a}}\hspace{1em}
	\subfloat[$D(\omega)$]{\includegraphics[scale=1]{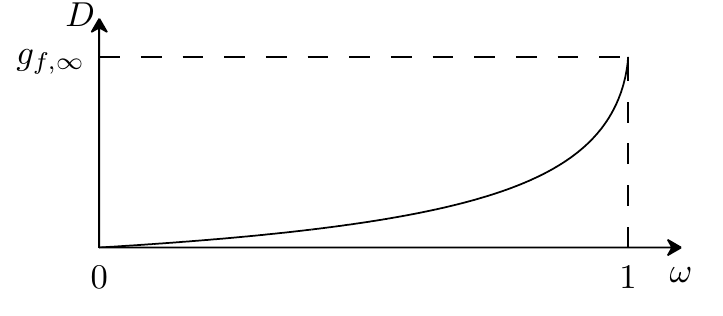}\label{Sect:A:Fig:1b}}
	\caption{A sketch of the damage variable~$\omega$ as a function of the strain~$\varepsilon$ corresponding to the exponential softening law defined in Eq.~\eqref{Examples:Eq:2} (cf. also Fig.~\ref{Sect:Examples:Fig:1}), and corresponding dissipation function~$D(\omega)$.}
	\label{Sect:A:Fig:1}
\end{figure}
\begin{equation}
\widehat{\omega} = g(\widehat{\varepsilon}) = 1-\frac{\varepsilon_0}{\widehat{\varepsilon}}\exp\left(-\frac{\widehat{\varepsilon}-\varepsilon_0}{\varepsilon_f}\right), \quad \varepsilon_0 \leq \widehat{\varepsilon},
\label{Examples:Eq:3}
\end{equation}
and is shown in Fig.~\ref{Sect:A:Fig:1a}. Rewriting Eq.~\eqref{Examples:Eq:3} as
\begin{equation}
\frac{\varepsilon_0 \exp(\varepsilon_0/\varepsilon_f)}{(1-\widehat{\omega})\varepsilon_f} = \frac{\widehat{\varepsilon}}{\varepsilon_f} \exp\left( \frac{\widehat{\varepsilon}}{\varepsilon_f} \right),
\label{Sect:A:Eq:1}
\end{equation}
we can cast its inversion in terms of the Lambert transcendental~$W$ function (recall the defining equation~$W(x) e^{W(x)} = x$, cf.~\cite{LambertFunc}, Eq.~(1.5))
\begin{equation}
\widehat{\varepsilon} = g^{-1}(\widehat{\omega}) = \varepsilon_f W\left( \frac{\varepsilon_0 \exp(\varepsilon_0/\varepsilon_f)}{(1-\widehat{\omega})\varepsilon_f} \right), \quad \widehat{\omega} \in [0,1],
\label{Sect:A:Eq:2}
\end{equation}
which is the counterpart to Eq.~\eqref{damageInversion}. Upon substituting this inversion in Eq.~\eqref{SubSect:FullSol:Eq:9}, we end up with the following integral
\begin{equation}
D(\widehat{\omega}) = C_1 \int_{0}^{\widehat{\omega}}	
\left[W\left( \frac{C_2}{1-\eta} \right)\right]^2
\,\mathrm{d}\eta,
\label{Sect:A:Eq:3}
\end{equation}
where
\begin{equation}
C_1 = \frac{EAr_0\varepsilon_f^2}{2},\quad
C_2 = \frac{\varepsilon_0 \exp(\varepsilon_0/\varepsilon_f)}{\varepsilon_f},
\label{Sect:A:Eq:4}
\end{equation}
have been introduced for brevity. Let us further denote~$w = W(C_2/(1-\eta))$ and write (by the defining equation of the Lambert function)
\begin{equation}
w e^w = \frac{C_2}{1-\eta},
\label{Sect:A:Eq:5}
\end{equation}
which can be differentiated on both sides to yield
\begin{equation}
(1 + w)e^w\,\mathrm{d}w = \frac{1}{C_2}\left( \frac{C_2}{1-\eta} \right)^2\,\mathrm{d}\eta.
\label{Sect:A:Eq:6}
\end{equation}
Employing Eq.~\eqref{Sect:A:Eq:5} in Eq.~\eqref{Sect:A:Eq:6}, we obtain
\begin{equation}
\mathrm{d}\eta = \frac{(1 + w)C_2}{w^2}e^{-w}\,\mathrm{d}w.
\label{Sect:A:Eq:7}
\end{equation}
Now, a change of variables in the integral~\eqref{Sect:A:Eq:3} according to Eq.~\eqref{Sect:A:Eq:5} can be carried out, providing us with
\begin{equation}
D(\widehat{\omega}) = C_1 \int_{0}^{\widehat{\omega}} w^2\,\mathrm{d}\eta 
= C_1 C_2 \int_{W(C_2)}^{W\left(\frac{C_2}{1 - \widehat{\omega}}\right)} (1 + w)e^{-w}\,\mathrm{d}w = - C_1 C_2 \, e^{-w}(2 + w) \biggr|_{W(C_2)}^{W\left(\frac{C_2}{1 - \widehat{\omega}}\right)}.
\label{Sect:A:Eq:8}
\end{equation}
The relation on the right hand side of Eq.~\eqref{Sect:A:Eq:8} can be expanded as
\begin{multline}
D(\widehat{\omega}) = \\
= \frac{EAr_0\varepsilon_f^2}{2}\left\{
\frac{\varepsilon_0}{\varepsilon_f}\left( 2 + \frac{\varepsilon_0}{\varepsilon_f} \right) -
(1-\widehat{\omega})W\left( \frac{\varepsilon_0\exp(\varepsilon_0/\varepsilon_f)}{\varepsilon_f(1-\widehat{\omega})} \right)
\left[ 2 + W\left( \frac{\varepsilon_0\exp(\varepsilon_0/\varepsilon_f)}{\varepsilon_f(1-\widehat{\omega})}\right) \right]
\right\}
\label{Sect:A:Eq:9}
\end{multline}
see also Fig.~\ref{Sect:A:Fig:1b} where a sketch of~$D(\widehat{\omega})$ is shown. The energy dissipated by the complete failure process then reads
\begin{equation}
g_{f,\infty} = \lim_{\widehat{\omega} \rightarrow 1} D(\widehat{\omega}) = EAr_0\varepsilon_0\left(\frac{\varepsilon_0}{2}+\varepsilon_f\right),
\label{Sect:A:Eq:10}
\end{equation}
which can be verified by integrating the area under the curve in Fig.~\ref{Sect:Examples:Fig:1}.
%
%
\section{Explicit Forms of Gradients and Hessians}
\label{Sect:B}
The first and second derivatives, i.e. the gradients and Hessians, of the incremental energy~$\Pi_\mathrm{red}^k$ with respect to kinematic variable~$\widehat{\bs{r}}$ are presented in this appendix. The internal force associated with atom~$\alpha$, $\bs{f}^\alpha_\mathrm{int}\in\mathbb{R}^{2\, n_\mathrm{ato}}$, is expressed as
\begin{multline}
\bs{f}^\alpha_{\mathrm{int},\gamma}(\widehat{\bs{r}})= 
\frac{\partial\widehat{\pi}^k_{\mathrm{red},\alpha}(\widehat{\bs{r}};\bs{q}(t_{k-1}))}{\partial\widehat{\bs{r}}^\gamma} =
\frac{1}{2}\sum_{\beta\in B_\alpha}\frac{\partial \left[(1-\mathring{\omega}^{\alpha\beta})\phi^{\alpha\beta}(\widehat{r}^{\alpha\beta}_+) + \phi^{\alpha\beta}(\widehat{r}^{\alpha\beta}_-) \right]}{\partial\widehat{\bs{r}}^{\gamma}} = \\
=\left\{
\begin{aligned}
&\frac{1}{2}\sum_{\beta\in  B_{\alpha}}(1-\mathring{\omega}^{\alpha\beta})\bs{f}^{\alpha\beta}(\delta^{\beta\gamma}-\delta^{\alpha\gamma}) && \mbox{if}\ \widehat{r}^{\alpha\beta} \geq r^{\alpha\beta}_0 \\
&\frac{1}{2}\sum_{\beta\in  B_{\alpha}}\bs{f}^{\alpha\beta}(\delta^{\beta\gamma}-\delta^{\alpha\gamma}) && \mbox{if}\ \widehat{r}^{\alpha\beta}<r^{\alpha\beta}_0, \\
\end{aligned}
\right. \quad \gamma=1,\dots,n_\mathrm{ato},
\label{Sect:B:Eq:1}
\end{multline}
where~$\widehat{\pi}^k_{\mathrm{red},\alpha}$ denotes the reduced incremental site energy with condensed internal variables in analogy to~$\Pi^k_\mathrm{red}$ defined in~\eqref{Sect:Sol:Eq:1} (see also Eqs.~\eqref{SubSect:DissLatt:Eq:7b}, \eqref{SubSect:FullSol:Eq:3}~-- \eqref{SubSect:FullSol:Eq:5}, and~\eqref{SubSect:QCSol:Eq:4}), and where the interatomic force reads
\begin{equation}
\bs{f}^{\alpha\beta} = \phi'(\widehat{r}^{\alpha\beta})\frac{\widehat{\bs{r}}^{\alpha\beta}}{\widehat{r}^{\alpha\beta}}.
\label{Sect:B:Eq:1a}
\end{equation}
The global force is then expressed as
\begin{equation}
\bs{f}(\widehat{\bs{r}}) = 
-\bs{f}_\mathrm{ext}(t_k)+
\sum_{\alpha=1}^{n_\mathrm{ato}}\bs{f}^\alpha_\mathrm{int}(\widehat{\bs{r}}).
\label{Sect:B:Eq:Eq:2}
\end{equation}
The stiffness matrix associated with an atom site~$\alpha$, $\bs{K}^\alpha\in\mathbb{R}^{2\, n_\mathrm{ato}\times 2\, n_\mathrm{ato}}$, reads
\begin{multline}
\bs{K}^{\alpha}_{\gamma\delta}(\widehat{\bs{r}}) = \frac{\partial^2\widehat{\pi}_\mathrm{red,\alpha}^k(\widehat{\bs{r}};\bs{q}(t_{k-1}))}{\partial\widehat{\bs{r}}^\gamma\partial\widehat{\bs{r}}^\delta}=
\frac{1}{2}\sum_{\beta\in B_\alpha}\frac{\partial^2 \left[ (1-\mathring{\omega}^{\alpha\beta})\phi^{\alpha\beta}(\widehat{r}^{\alpha\beta}_+) + \phi^{\alpha\beta}(\widehat{r}^{\alpha\beta}_-) \right]}{\partial\widehat{\bs{r}}^{\gamma}\partial\widehat{\bs{r}}^{\delta}}=\\
 =
\left\{
\begin{aligned}
&\frac{1}{2}\sum_{\beta\in B_\alpha}(1-\mathring{\omega}^{\alpha\beta})\bs{K}^{\alpha\beta}(\delta^{\beta\gamma}-\delta^{\alpha\gamma})(\delta^{\beta\delta}-\delta^{\alpha\delta}) &&\mbox{if}\ \widehat{r}^{\alpha\beta} \geq r^{\alpha\beta}_0 \\
&\frac{1}{2}\sum_{\beta\in B_\alpha}\bs{K}^{\alpha\beta}(\delta^{\beta\gamma}-\delta^{\alpha\gamma})(\delta^{\beta\delta}-\delta^{\alpha\delta}) &&\mbox{if}\ \widehat{r}^{\alpha\beta}<r^{\alpha\beta}_0, \\
\end{aligned}\right.
\quad \gamma,\delta=1,\dots,n_\mathrm{ato},
\label{Sect:B:Eq:3}
\end{multline}
where the interaction Hessian reads
\begin{equation}
\bs{K}^{\alpha\beta} = \left[\frac{\phi'(\widehat{r}^{\alpha\beta})}{\widehat{r}^{\alpha\beta}}\bs{I}_2+\left(\frac{\phi''(\widehat{r}^{\alpha\beta})}{(\widehat{r}^{\alpha\beta})^2}-\frac{\phi'(\widehat{r}^{\alpha\beta})}{(\widehat{r}^{\alpha\beta})^3}\right)\widehat{\bs{r}}^{\alpha\beta}\otimes\widehat{\bs{r}}^{\alpha\beta}\right].
\end{equation}
The global stiffness is then expressed as
\begin{equation}
\bs{K}(\widehat{\bs{r}})=\sum_{\alpha=1}^{n_\mathrm{ato}}\bs{K}^\alpha(\widehat{\bs{r}}).
\label{Sect:B:Eq:4}
\end{equation}
Above, we have used the relation
\begin{equation}
\frac{\partial\widehat{r}^{\alpha\beta}}{\partial\widehat{r}_m^\gamma}=\frac{\widehat{r}_m^{\alpha\beta}}{\widehat{r}^{\alpha\beta}}(\delta^{\beta\gamma}-\delta^{\alpha\gamma}),\ m=1,2,
\label{Sect:B:Eq:5}
\end{equation}
and for brevity, as in~\ref{Sect:A}, have denoted
\begin{equation}
\phi'(\widehat{r})=\frac{\mathrm{d}}{\mathrm{d}\widehat{r}}\phi^{\alpha\beta}(\widehat{r}),\quad \phi''(\widehat{r})=\frac{\mathrm{d}^2}{\mathrm{d}\widehat{r}^2}\phi^{\alpha\beta}(\widehat{r}).
\label{Sect:B:Eq:6}
\end{equation}
The symbol~$\bs{I}_2 \in \mathbb{R}^{2 \times 2}, (\bs{I}_2)_{mn} = \delta_{mn}$, denotes the identity matrix, $\delta_{mn}$ denotes the Kronecker-delta product with respect to spatial coordinates, $m, n$ indices relate to spatial dimensions, $\alpha,\beta$ relate to atoms, $\delta^{\alpha\beta}$ denotes the Kronecker-delta product with respect to atoms, and~$\bs{a}\otimes\bs{b}=a_mb_n$ is the tensor product of vectors~$\bs{a}$ and~$\bs{b}$.
%
%
\section*{Acknowledgements}
Financial support of this work from the Czech Science Foundation (GA\v{C}R) under project No.~14-00420S is gratefully acknowledged.

%
%


\end{document}